\RequirePackage{fix-cm}
\documentclass[smallextended]{svjour3}       
\smartqed  
\usepackage{graphicx}
\usepackage{natbib}
\usepackage{color}
\usepackage{times}
\usepackage{latexsym}
\usepackage{enumerate}
\usepackage{epsfig}
\usepackage{amsmath}
\usepackage{amssymb}
\usepackage{amsfonts}
\usepackage[english]{babel}
\usepackage{xcolor}

\def\bb{\mathbf{b}}

\def\bx{\mathbf{x}}

\def\bd{\mathbf{d}}

\def\bv{\mathbf{v}}
\def\bu{\mathbf{u}}

\def\bX{\mathbf{X}}

\def\bZ{\mathbf{Z}}
\def\b0{\mathbf{0}}
\def\b1{\mathbf{1}}

\def\cX{\mathcal{X}}
\def\cY{\mathcal{Y}}

\def\mR{\mathbb{R}}

\def\bmu{\mbox{\boldmath $\mu$}}

\def\bpsi{\mbox{\boldmath $\psi$}}

\def\bbeta{\mbox{\boldmath $\beta$}}

\def\bxi{\mbox{\boldmath $\xi$}}
\def\btheta{\mbox{\boldmath $\theta$}}

\def\bSigma{\mbox{\boldmath $\Sigma$}}

\newtheorem{mytheorem}{Proposition}
\numberwithin{mytheorem}{subsection} 

\begin{document}

\title{Robust estimation of mixtures of regressions with random covariates, via trimming and constraints
}

\titlerunning{Robust estimation for mixture of regression models}        
 \author{L.A. Garc\'{i}a-Escudero \and A. Gordaliza  \and \\F. Greselin \and S. Ingrassia \and A. Mayo-Iscar
 }

\authorrunning{L.A. Garc\'{i}a-Escudero et al.} 

\institute{L.A. Garc\'{i}a-Escudero, A. Gordaliza, A. Mayo-Iscar  \at
             Department of Statistics and  Operations Research and IMUVA, University of Valladolid, Valladolid, Spain 
              \email{lagarcia@eio.uva.es; alfonsog@eio.uva.es, agustinm@eio.uva.es}           
       \and
          F. Greselin  \at
              Department of Statistics and Quantitative Methods, Milano-Bicocca University, Milano,  Italy
              \email{francesca.greselin@unimib.it}           
      \and
          S. Ingrassia   \at
              Department of Economics and Business, University of Catania, Catania, Italy
              \email{s.ingrassia@unict.it}           
}


\date{Received: date / Accepted: date}

\maketitle

\begin{abstract}

A robust estimator for a wide family of mixtures of linear regression is presented. Robustness is based on the joint adoption of the Cluster Weighted Model and of an estimator based on trimming and restrictions. The selected model provides the conditional distribution of the response for each group, as in mixtures of regression, and further supplies local distributions for the explanatory variables. A novel version of the restrictions has been devised, under this model, for separately controlling the two sources of variability identified in it. This proposal avoids singularities in the log-likelihood, caused by approximate Òlocal collinearityÓ in the explanatory variables or Òlocal exact fitsÓ in regressions, and reduces the occurrence of spurious local maximizers. In a natural way, due to the interaction between the  model and the estimator, the procedure is able to resist the harmful influence of ÓbadÓ leverage points along the estimation of the mixture of regressions, which is still an open issue in the literature.
The given methodology defines a well-posed statistical problem, whose estimator exists and is consistent to the corresponding solution of the population optimum, under widely general conditions. A feasible EM algorithm has also been  provided to obtain the corresponding estimation.
Many simulated examples and two real datasets have been chosen to show the ability of the procedure, on the one hand, to detect anomalous data, and,  on the other hand, to identify the real cluster regressions without the influence of contamination.

\end{abstract}

\section{Introduction}
\label{intro}
Mixture models provide a quite flexible approach to statistical
modeling of a wide variety of random phenomena, whenever we can
reasonably suppose that the observations arise from unobserved groups
in the population. Under this general framework, the present paper
provides a new proposal in the family of finite mixtures of robust
regressions \citep{DeSa:Cron:Amax:1988,deVR89}.

Assume we are provided with two quantitative random variables $\bX$
and $Y$: $\bX$ is a vector of \textit{explanatory} variables, $Y$ is
a \textit{response} or \textit{outcome} variable, and  the
dependence between $Y$ and $\bX$ may vary among the different
underlying groups. By adopting the cluster-weighted approach, 
we  allow  different scatter structures in each group, both in
the marginal distribution of $\bX$ and in the conditional
distribution of $Y |\bX = \bx$, as it is required 
by many observed
dataset. The Cluster Weighted Model (CWM), introduced in
\citet{Gers:Nonl:1997}, decomposes the joint p.d.f. of $(\bX,Y)$ in
each component of the mixture as the product of the marginal and the
conditional distributions.

Due to its very definition, the CWM estimator is able to take into account 
different distributions for the explanatory variables across groups, so overcoming an intrinsic limitation of mixtures of regression, where they are implicitly assumed equally distributed. 
However, due to the possible presence of
contaminating data (background noise, pointwise contamination, unexpected minority patterns, etc.) a
small fraction of outliers could severely affect the model fitting. Among the available standard
techniques in robust estimation, those based on removing part of the data - and called
\textit{impartial trimming procedures} - present a good performance, often being an obligatory
benchmark to compare new estimators. Successful robust procedures of this kind are, for instance, the LTS for  regression models \citep{RouL87},
the trimmed k-means \citep{CueG97}, the TCLUST for
clustering \citep{GarG08}, and the robust clusterwise linear regression models \citep{GarS10}. Here, in
the framework of mixtures of regressions, denoting by $\bx$ and $y$ the realizations of $\bX$ and $Y$, standard
diagnostic tools can easily identify outliers on $y$ that fall in the range of values of $\bx$, while the detection of
outliers on both $\bx$ and $y$, that may act as ÒbadÓ leverage points, is much more problematic. 
Many trimming approaches are effective for the first
type of outliers, but they fail when dealing with bad leverage
points. In this paper, we exploit the CWM nice feature of
modeling the $\bX$ marginal distribution, to detect dangerous
outliers on $\bx$. At the same time, we also use the regression structure among $\bX$ and $Y$
to deal with outliers on $y$. In this way, by robustifying the CWM estimation,  we can simultaneously handle  both type of
outliers with the same formal approach. As usual when using trimming, only the total
fraction of discarded observations must be fixed in advance.

A further issue with ML estimation for CWMs is the
unboundedness of the log-likelihood function, a well-known aspect
pointed out in \cite{Day:Esti:1969} for  Gaussian mixtures. To overcome this drawback, \cite{Hath:Acon:1985} 
introduced the use of constrained variance estimation in univariate mixture modeling. These restrictions
 have been extended to the multivariate case in
different ways by \cite{McLaPe:2000}, \cite{IngR07} and  \cite{GarG08}. By adopting restrictions also for CWM, we arrive at setting a well-posed optimization problem.
Additionally, a restricted approach not only avoids singularities,
it also discards non-interesting local maximizers of the objective function \citep{GarG:ADAC13}. We will discuss in
detail how approximate Òlocal collinearityÓ in the explanatory variables, and approximate Òlocal exact
fitsÓ in the regressions may cause, indeed, serious troubles in CWMs. 

The above considerations give rise to the robust estimation of the trimmed Cluster Weighted Restricted Model (trimmed CWRM) presented hereafter.
 It includes an original application of the constraints, which takes into
account the specific features of CWM and controls the relative variability between components for the sources of variability in the model corresponding to:  i) the explanatory variables, and ii)  the
regression errors. The CWM, endowed with restrictions and trimming, becomes a very
competitive robust estimator for mixtures of multiple regression, with optimal statistical properties.

We have organized the paper as follows. In Section \ref{sec:cwm_intro} we recall the main ideas about the CWM. In Section \ref{se3}  we present the trimmed CWRM, and introduce a feasible  algorithm for its practical implementation. Then, we state the central findings of the paper, i.e.  the existence and the strong consistency of the new estimator. Section \ref{se4} provides a discussion on the effects of
constraints and trimming, along with some illustrative examples. The
application of the proposed methodology to two real data sets is
shown in Section \ref{se5}. Finally, Section \ref{se6} contains some
concluding remarks and sketches future research. Proofs and  technical lemmas needed for our main results are relegated in the Appendix.

\section{Cluster Weighted Modeling}\label{sec:cwm_intro}

The Cluster Weighted Model (CWM) has been proposed in the context of
media technology, to build a digital violin with traditional inputs
and realistic sounds
\citep{Gers:Nonl:1997,Gersh+Schoner+Metois:cwmNature:1999}; in
\citet{Wedel:user:2000}.  CWMs are referred to as the family of
saturated mixture regression models. In
\citet{Ingr:Mino:Vitt:Loca:2012}, CWMs have been reformulated in a
statistical  setting showing that they are a general and flexible
family of mixture models. In fact,  \citet{Ingr:Mino:Vitt:Loca:2012}
show that Gaussian CWM includes, as special cases, finite mixtures
of distributions and finite Mixtures of Regression models.

Let $(\bX,Y)$ be a pair  of random variables, namely a vector of
covariates $\bX$ and a response variable $Y$ defined on $\Omega$
with values in   $\cX \times \cY \subseteq \mR^d \times \mR$ and
$\{(\bx_i,y_i)\}_{i=1}^n$ represents a i.i.d. random sample of size
$n$, drawn from $(\bX,Y)$. Let $p(\bx,y)$ denote the joint density
of $\left(\bX,Y\right)$, and suppose that $\Omega$ can be
partitioned into $G$ groups, say $\Omega_1, \ldots,\Omega_G$. CWMs
are mixture models having density
 of the form
\begin{equation}
p (\bx, y; \btheta )=\sum^{G}_{g=1} p(y|\bx; \bxi_g ) p(\bx;
\bpsi_g) \pi_g , \label{eq:CWM base}
\end{equation}
where $p(y|\bx; \bxi_g )$ is the conditional density of  $Y$ given
$\bx$  in $\Omega_g$ (depending on some parameter $\bxi_g$), $p(\bx;
\bpsi_g)$ is the marginal density of $\bX$ in $\Omega_g$ (depending
on some parameter $\bpsi_g$) and $\pi_g$ is the weight of $\Omega_g$
in the mixture (with $\pi_g > 0$ and $\sum_{g=1}^G \pi_g = 1$).
Furthermore, we assume that in each group $\Omega_g$, the
conditional expectation of $Y$ given $\bX=\bx$, is a function
$m(\cdot)$ of $\bx$ depending on some parameters $\bbeta_g$, that is
$E(Y|\bx, \Omega_g )= m(\bx; \bbeta_g)$.

In this work, we have focused on models of type \eqref{eq:CWM base}
with Gaussian components. Thus
 $p(\bx; \bpsi_g)= \phi_d (\bx;\bmu_g,\bSigma_g)$,
where $\phi_d (\cdot;\bmu_g,\bSigma_g)$ denotes the density of the
$d$-variate Gaussian distribution with mean vector $\bmu_g$ and
covariance matrix $\bSigma_g$. Moreover, we have assumed that the
conditional relationship between $Y$ and $\bx$ in the $g$-th group
can be written  as $Y=  \bb_g' \bx+b_g^0+\varepsilon_g$ where
$\varepsilon_g\sim N(0, \sigma_g^2)$. Hence, $\bX|\Omega_g \sim
N_d(\bmu_g, \bSigma_g)$ and $Y |\bx, \Omega_g \sim N(\bb'_g
\bx+b_g^0,\sigma_g^2)$, so that
 model \eqref{eq:CWM base} specializes to:
\begin{equation}
p(\bx, y;\btheta)=\sum_{g=1}^G \phi(y;\bb'_g
\bx+b_{g}^0,\sigma_g)\phi_d (\bx;\bmu_g,\bSigma_g)\pi_g,
\label{eq:GausslinCWM}
\end{equation}
which  defines the {\em linear Gaussian CWM}.  We notice here that
definition  (\ref{eq:GausslinCWM}) corresponds to a mixture of
regressions, with weights $\phi_d (\bx;\bmu_g,\bSigma_g)\pi_g$
depending also on the covariate distributions in each component $g$
for $g=1,\ldots,G$. Finally, in the framework of model-based
clustering, each unit is assigned to one group, based on the maximum
a posteriori probability. The consideration of
(\ref{eq:GausslinCWM}) yields to the use of (log-)likelihood target
function to be maximized as
\begin{equation}\label{eq:targetfunction}
\sum_{i=1}^n \log\left[\sum_{g=1}^G \phi(y_i;\bb'_g \bx_i
+b_{g}^0,\sigma_g^2)\phi_d (\bx_i ;\bmu_g,\bSigma_g)\pi_g \right].
\end{equation}

For sake of simplicity, we will later use the notation
$$D_g(\bx,y;\btheta)=\phi(y;\bb'_g \bx +b_{g}^0,\sigma_g^2) \phi_d
(\bx ;\bmu_g,\bSigma_g)\pi_g$$ and $D(\bx,y;\btheta)\allowbreak
=\sum_{g=1}^G D_g(\bx,y;\btheta)$, where the set of all parameters
of the model is denoted by $\btheta$, and, such that
(\ref{eq:targetfunction}) is simply rewritten as
$\sum_{i=1}^n\log[D(\bx_i,y_i;\btheta)]$. Additionally, the linear
Gaussian CWM will be many times simply referred to as CWM.

\subsection{Two problems about CWM}
\begin{figure}[ht]
\centering
\includegraphics[width=11.5cm]{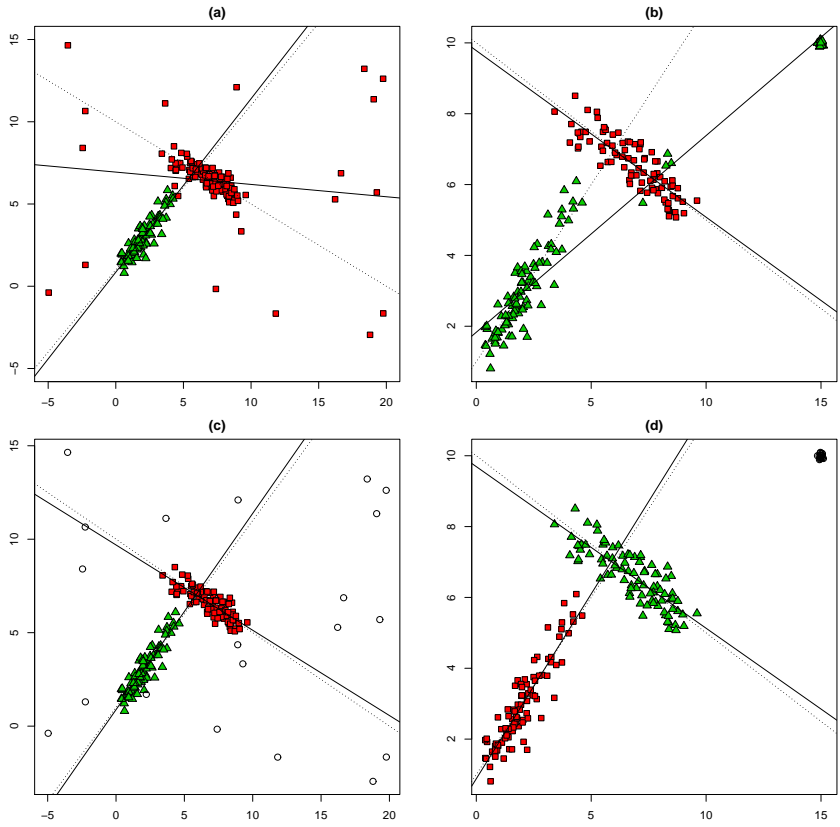}
\caption{ {\em Simdata1}: (a) original data plus  background noise
and CWM fitted; (b) original data plus pointwise contamination and
CWM fitted; (c) and (d) show the fitted trimmed CWRMs with
$\alpha=0.1$, $c_X=c_{\varepsilon}=20$ to these two data sets. The
dotted lines represent the true regression lines to be estimated and
black circles are the trimmed observations (here and in all the
figures).} \label{fig:simdata1}
\end{figure}

 The  estimation of the (linear Gaussian) CWM  suffers from a
serious lack of robustness, like it happens when using many other
models based on normal assumptions and fitted through ML estimators
\citep[see, e.g.,][]{Huber:1981}. It is very important to be aware
of this issue, due to the common presence of noise sources in data.
To illustrate this problem, a simulated data set  of $n=180$ units
(referred to as {\em Simdata1} hereafter), has been generated from
the CWM  with $G=2$ and 90 observations from each component. Then we
added 20 contaminating observations as either background noise, see
Figure \ref{fig:simdata1}(a), or pointwise contamination around the
point $(15,20)$, see Figure \ref{fig:simdata1}(b). The true
underlying regression lines (prior to contamination) are represented
with dotted lines, and we can see the dangerous effects of outliers
on model fitting for the standard CWM.

Another important issue concerns the unboundedness of the target
function in (\ref{eq:targetfunction}) when no constraints are imposed
on the scatter parameters. In this case, the defining problem is
ill-posed because the 
loglikelihood in \eqref{eq:targetfunction} tends to $\infty$ when
either $\bmu_g=\bx_i$ and $|\bSigma_g| \rightarrow 0$ or $y_i=\bb'_g
\bx_i +b_{g}^0$ and $\sigma_g^2
\rightarrow 0$. Moreover, as a trivial consequence, 
the EM algorithms often applied to fit a CWM can be trapped into
non-interesting local maximizers, called ``spurious" solutions, and
the result of the
EM algorithm strongly depends on its initialization. 

Spurious solutions may be due to very localized patterns in the
explanatory variables, as shown in Figure \ref{fig:simdata2}(a), by
considering a second simulated data set ({\em Simdata2}). Here, data
concern $n=200$ observations and $d=2$ explanatory variables. The
dataset has been built as follows: two sets of 90 observations for
the explanatory variable $\bX$ has been drawn from two bivariate
normal distributions,  centered at $(2,2)$ and $(4,4)$,
respectively. Then, 20 almost collinear observations have been added
to the sample, close to the second component. The values for the
response variable $Y$ have been generated by using the same linear
function (for both components) with equally distributed error terms.
We can see in Figure \ref{fig:simdata2}(a) that the standard fit of
the CWM yields to the determination of a first spurious component
with the 20 almost collinear observations and a second component
joining together the two  groups, with $90\%$ of the observations.

Sometimes spurious solutions may be also due to localized patterns
of observations, where an approximate ``exact fit" for a small
number of observations can be obtained. Figure \ref{fig:simdata3}
shows a third simulated data set ({\em Simdata3}) with $n=200$
observations, where 196 of them have been generated from a CWM with
$G=2$ components (98 observations from each component). A very small
fraction of almost collinear units (only 4 observations) on the
$(\bX,Y)$ variables have been added, with a roughly equal value
(around $0$) for the response variable. These values, for instance,
could be due to a bad performance of the tool used to measure the
response variable. It may be seen that a fitted component including
only these almost collinear observations could arise, along the EM
estimation, because a small value of one of the $\sigma_g^2$
parameters yields to higher values of the log-likelihood. Then, the
two main linear structures accounting for $98\%$ of the data points
would be artificially joined together.

To overcome the previous issues, in the next section we propose a
robust methodology by incorporating trimming and constraints to the
CWM.

\section{Trimmed Cluster Weighted Restricted Modeling}\label{se3}

\subsection{Problem statement}\label{se3_1}

For a given sample of $n$ observations, the trimmed CWRM methodology
is based on the maximization of the following log-likelihood
function
\begin{equation}\label{d1}
    \sum_{i=1}^n z(\bx_i, y_i) \log\left[\sum_{g=1}^G \phi(y_i ;\bb'_g \bx_i +b_{g}^0,\sigma_g^2)\phi_d (\bx_i ;\bmu_g,\bSigma_g)\pi_g \right],
\end{equation}
where $z(\cdot,\cdot)$ is a 0-1 trimming indicator function that
tell us whether observation 
$(\bx_i,y_i)$ is trimmed off ($z(\bx_i, y_i)$=0), or not ($z(\bx_i,
y_i)$=1). A fixed fraction $\alpha$ of observations can be
unassigned by setting $\sum_{i=1}^nz(\bx_i, y_i)=[n(1-\alpha)]$.
Hence the parameter $\alpha$ denotes the trimming level. Analogous
approaches based on trimmed mixture likelihoods can be found in
\cite{NeyF07}, \cite{GalR09} and \cite{GarG:ADAC13}.

Moreover, we introduce two further constraints on the maximization
in (\ref{d1}). The first one 
concerns  the set of eigenvalues
$\{\lambda_l(\bSigma_g)\}_{l=1,...,d}$ of the scatter matrices
$\bSigma_g$ by imposing 
\begin{equation}
    \lambda_{l_1}(\bSigma_{g_1})\leq c_X
    \lambda_{l_2}(\bSigma_{g_2})\quad \quad \text{ for every }1 \leq l_1  \neq l_2 \leq
    d\text{ and }1 \leq g_1 \neq g_2 \leq G. \label{c_X}
\end{equation}

The second constraint refers to the  variances $\sigma_g^2$ of
the regression error terms, by requiring 
\begin{equation}
    \sigma_{g_1}^2\leq c_{\varepsilon}
    \sigma_{g_2}^2 \quad \quad \text{ for every }1 \leq g_1 \neq g_2 \leq  G. \label{c_eps}
\end{equation}
The constants $c_X$ and $c_{\varepsilon}$, in \eqref{c_X} and
\eqref{c_eps} respectively, are finite (not necessarily equal) real
numbers, such that $c_X \geq 1, c_{\varepsilon} \geq 1 $. They
automatically guarantee that we are avoiding the $|\bSigma_g|
\rightarrow 0$ and $\sigma_g^2 \rightarrow 0$ cases.  These
constraints are an extension to CWMs of those introduced in
\citet{IngR07}, \citet{GarG08} and \cite{Gres:Ingr:Cons:2010} and go
back to \citet{Hath:Acon:1985}. The main difference is the
asymmetric and different treatment  given by the constraints, when
modeling the marginal distribution $\bX$ or when modeling the
regression error terms, providing high flexibility to the model.

Let us consider now the effects of trimming in the two data sets
derived from \textit{Simdata1}. In Figure \ref{fig:simdata1}(c) and
(d) we can see that setting $\alpha=0.1$ allows to restore the true
structure of the data, by discarding the outlying observations, both
in the case of background noise and pointwise contamination. Hence,
trimming  modifies the ML estimation in such a way that it is no
more influenced by potential outliers and drives it far from the
previous bad results.

\begin{figure}[!h]
\centering
\includegraphics[width=6cm]{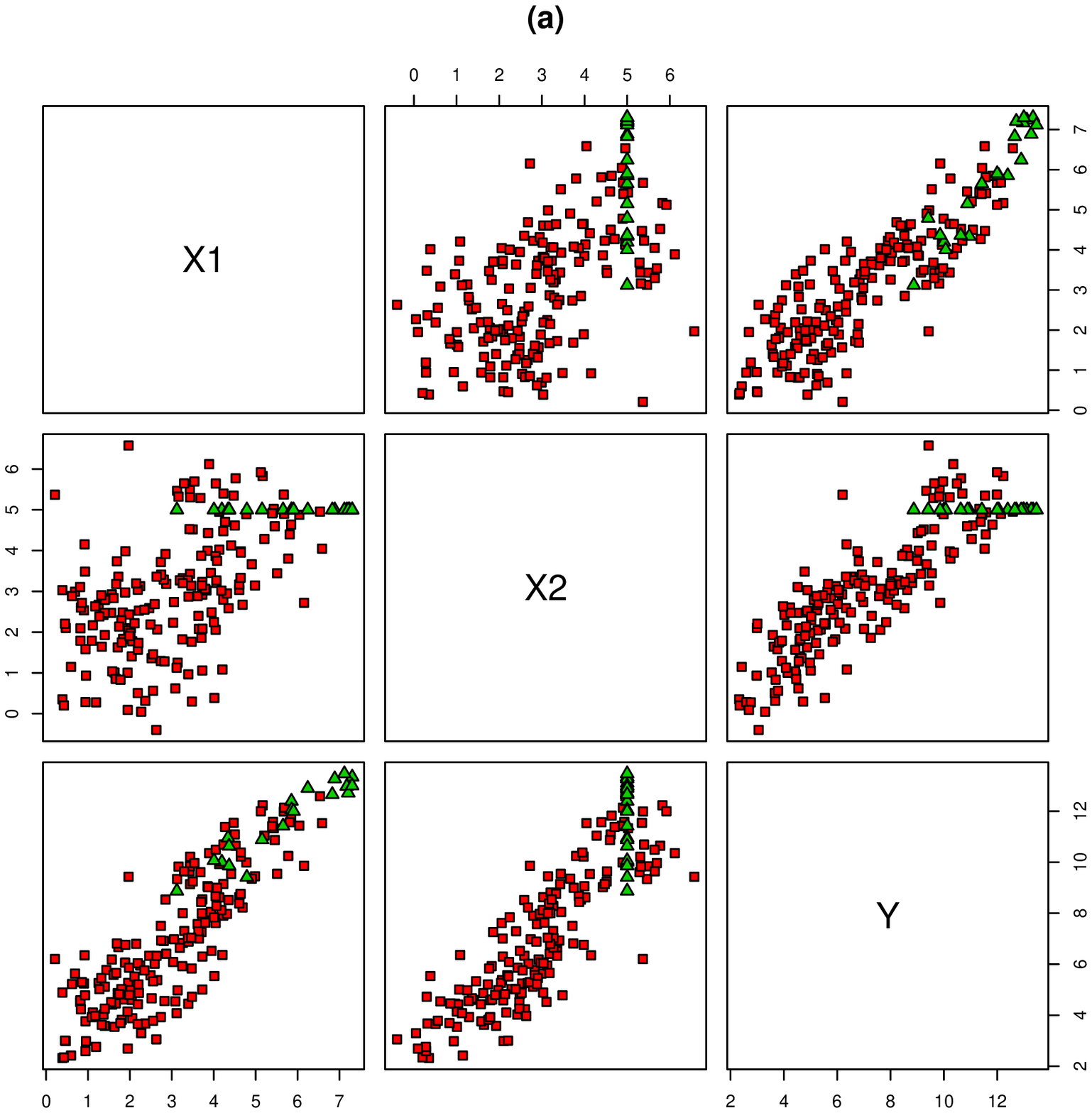} \includegraphics[width=6cm]{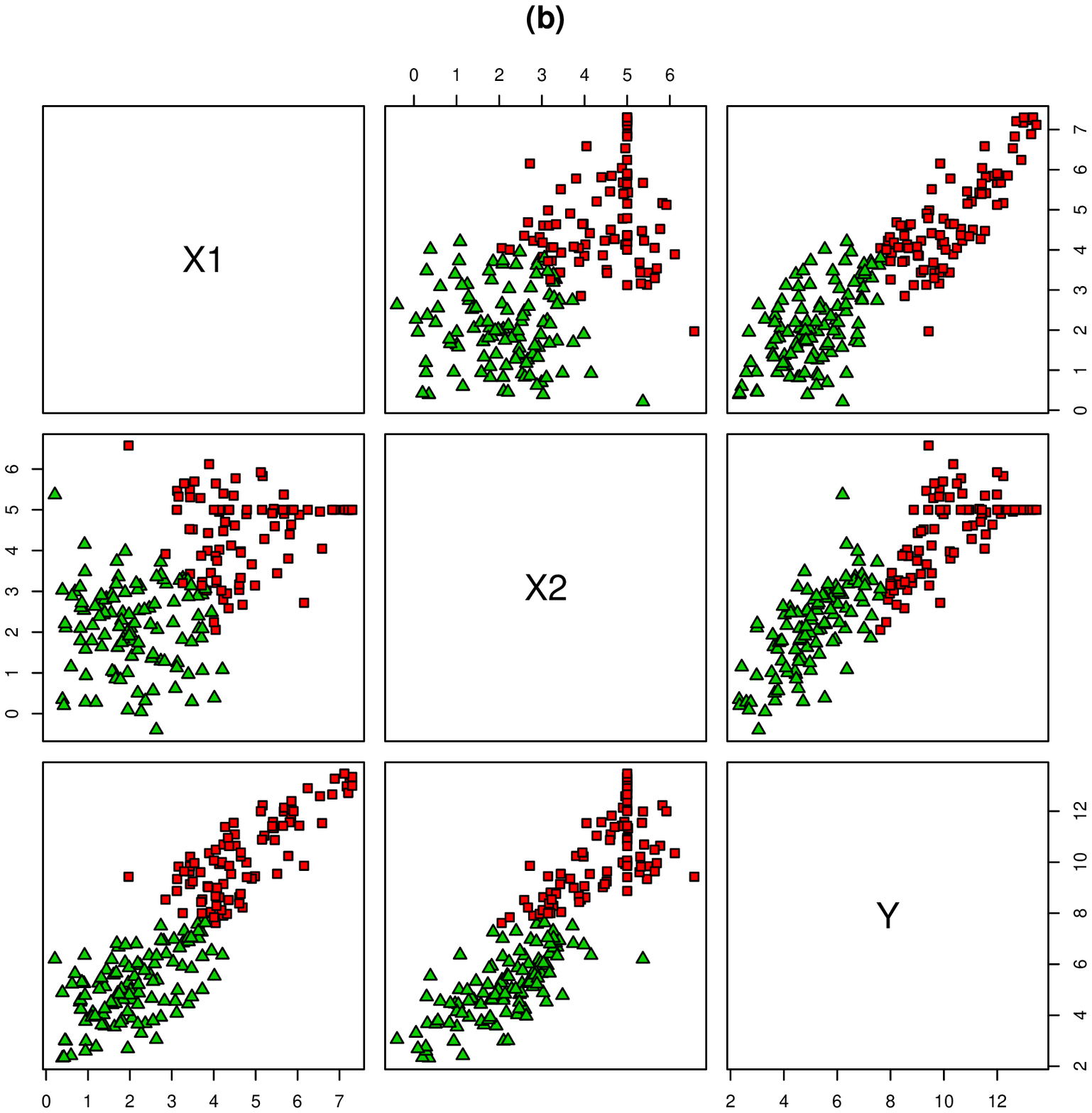}
\caption{{\em Simdata2}: Scatter plot matrix. (a) Almost collinear
observations in the explanatory variables which are found as a
cluster by CWM when $G=2$; (b) Results of fitting the trimmed CWRM
with $\alpha=0$, $c_X=c_{\varepsilon}=20$.}\label{fig:simdata2}
\end{figure}

Commenting the use of constraints, we can see how a moderate choice
of $c_X$ for  \textit{Simdata2} in Figure \ref{fig:simdata2}(b)
allows to correctly detect  the $G=2$ main groups and to avoid the
disturbing effect of the spurious patterns in the explanatory
variables.

\begin{figure}[!h]
\centering
\includegraphics[width=6cm]{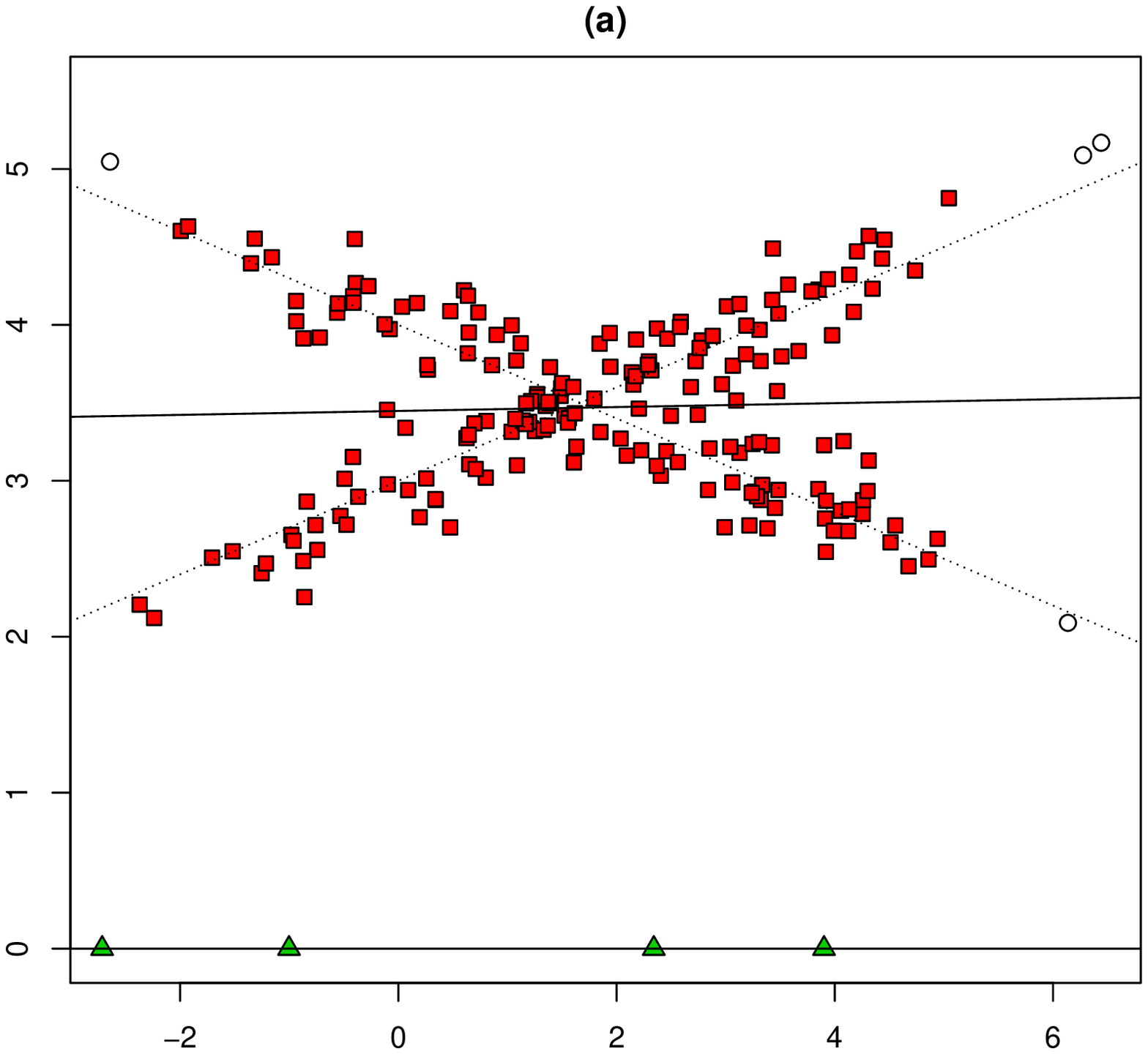} \includegraphics[width=6cm]{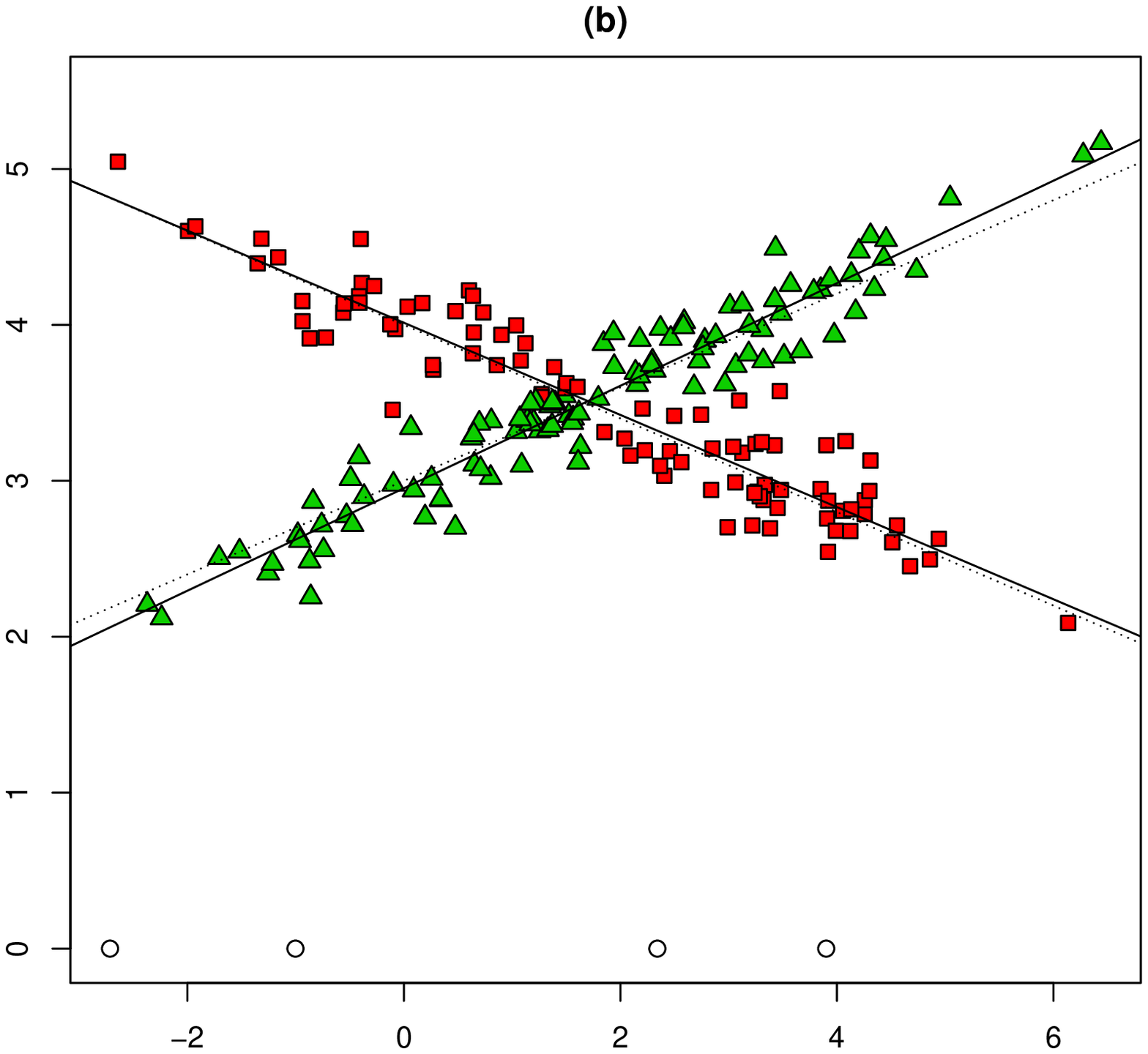}
\caption{{\em Simdata3:} (a) Results of the  trimmed CWRM fit with
$G=2$, $\alpha=0.02$ and $c_{\varepsilon}=10^{10}$ (almost
unrestricted) showing the detection of a spurious component due to
an approximate ``local exact fit" in one of the fitted regressions;
(b) results with $\alpha=0.02$,
$c_X=c_{\varepsilon}=20$.}\label{fig:simdata3}
\end{figure}

Additionally, we can see that a moderate choice of $c_{\varepsilon}$
for \textit{Simdata3} would also allow to correctly detect the $G=2$
main groups. Moreover, we can see in Figure \ref{fig:simdata3}(a)
how only considering $\alpha=0.02$ trimming level (trying to discard
the 4 outlying observations in \textit{Simdata3}) does not solve the
problem at all without the consideration of a moderate value of
$c_{\varepsilon}$.



A detailed discussion about the role played by $\alpha$, $c_X$ and
$c_{\varepsilon}$ is given in Section \ref{se4}.

\subsection{Theoretical results}\label{se3_2}

The problem stated in Section \ref{se3_1} admits a population
counterpart. Let $P=P_{(\bX,Y)}$ be the probability measure in
$\mathbb{R}^{d+1}$ induced by the joint distribution of the random
variables $\bX$ and $Y$ and let $E_P(\cdot)$ denote the expectation
with respect to $P$. 
Let $\Theta_{c_X,c_{\varepsilon}}$ denote
hereafter the set of all possible $\btheta$ which do satisfy
constraints (\ref{c_X}) and (\ref{c_eps}) for given constants $c_X$
and $c_{\varepsilon}$. With this notation, the population problem is
defined through the double maximization of $ E_P\big[\log
D(\bX,Y;\btheta)I_A(\bX,Y)\big] $ over all possible $\btheta \in
\Theta_{c_X,c_{\varepsilon}}$, and over all possible subsets
$A\subset \mathbb{R}^{d+1}$,  with $P[A]\geq 1-\alpha$. As usual,
$I_A(\cdot)$ denotes the indicator function of set $A$. We will see
that the optimal set $A$ can be determined directly from $\btheta$.
In more detail, fixed $\btheta$, and denoting by
$$
    R(\btheta,P)=\sup_u\big\{u:P[(\bX,Y):D(\bX,Y;\btheta)\geq u]\geq
1-\alpha\big\},
$$
then $A$ is given by $ A(\btheta)=
A(\btheta,P)=\{(\bx,y):D(\bx,y;\btheta)\geq R(\btheta,P)\}. $
Therefore, we reduce the population problem to that of maximizing
\begin{equation}\label{pop_pro}
    L(\btheta,P)=E_P\big[\log
    D(\bX,Y;\btheta)I_{A(\btheta)}(\bX,Y)\big],\text{ on }\btheta \in \Theta_{c_X,c_{\varepsilon}}
\end{equation}

Note that we recover the original sample problem introduced in
Section \ref{se3_1}, just by taking $P$ equal to the empirical
measure $P_n=\sum_{i=1}^n \delta_{\{(\bx_i,y_i)\}}$  and setting
 $z(\bx_i,y_i)=I_A(\bx_i,y_i)$ for the optimal
set $A$. The way that the optimal set $A$ is obtained from $\btheta$
will be also used in the C-steps of the algorithm to be presented in
Section \ref{se3_3}.

In this section, we present results guaranteeing the existence of
the solutions for both the sample and the population problem.
Moreover, we state the consistency  of the sample solution to the
population one. These results are derived under very mild
assumptions on the underlying distribution $P$. In fact, no moment
conditions are needed on $P$ and, thus, the proposed methodology can
be applied even to heavy-tailed distributions. We will only exclude
for $P$ some ``pathological" cases that are
clearly non appropriate, namely: 
\begin{quote}
    \textbf{(PR)} The support of $P$ is not concentrated on $G$
    regression hyperplanes and the support of $\bX$ is not concentrated
    in $G$ points in $\mathbb{R}^{d}$, after removing a probability mass equal
    to  $\alpha$,
\end{quote}
where we say that $S\subset \mathbb{R}^{d+1}$ is concentrated in a
``regression hyperplane" if an ``exact fit" property holds for some
$b^0$ and $\bb$ in such a way that $y=\bb' \bx+ b^0$ for all
$(\bx,y)\in S$. The previous condition holds for absolutely
continuous distribution $P$ as well as empirical measures $P_n$
obtained from absolutely continuous distributions when $n$ is large
enough.

\begin{mytheorem}
\label{p1} If \emph{(PR)} holds for $P$, then there exists
$\btheta\in \Theta_{c_X,c_{\varepsilon}}$ maximizing $L(\btheta,P)$.
\end{mytheorem}

The underlying distribution $P$ is typically unknown and we often
only rely on the result of a random sample from $P$. Let
$\widehat{\btheta}_n$ denote the solution of the sample problem for
a random sample of size $n$. If the population problem has a unique
solution $\btheta_0$, then the following property states that
$\widehat{\btheta}_n$ should be close to $\btheta_0$ when $n$ is
large.

\begin{mytheorem}
\label{p2} Assume that $P$ be an absolutely continuous distribution
with strictly positive density function satisfying \emph{(PR)} and
that $\btheta_0$ is the unique maximizer of $L(\btheta,P)$ for
$\btheta\in \Theta_{c_X,c_{\varepsilon}}$. If
$\{\widehat{\btheta}_n\}_{n=1}^{\infty}\subset
\Theta_{c_X,c_{\varepsilon}}$ is a sequence of maximizers of
\emph{(\ref{pop_pro})} when $P$ is replaced by the sequence of
empirical measures $\{P_n\}_{n=1}^{\infty}$, referred to a sequence
of i.i.d. samples from $P$, then $\widehat{\btheta}_n \rightarrow
\btheta_0$ almost surely.
\end{mytheorem}

Note that, apart from the (PR) condition, a uniqueness condition is
also needed to get  consistency. It is also important to note that
the parameters obtained by solving the maximization (\ref{pop_pro})
do not necessarily coincide with the parameters of the mixture
components appearing in the definition of the (uncontaminated) CWM.
However, we conjecture that these two different types of parameters
are ``close" each other whenever the contamination is not very
overlapped with the most interior regions of the mixture components
and when $\alpha$, $c_X$ and $c_{\varepsilon}$ are ``properly"
chosen. However, establishing results formalizing this idea is not
an easy task (as happens even in simpler clustering approaches).

Although the proofs of these theoretical results, given in the
Appendix, are related to previous works in \cite{GarG08} and
\cite{GarG:STCO13}, several specific technicalities must be sorted
out for the present case. In fact, these technicalities are far from
being straightforward and mainly have to do with how to deal with
the effect of ``local collinearities" in the regression
coefficients.

\subsection{Algorithm}\label{se3_3}

The constrained maximization of the trimmed log-likelihood in
(\ref{d1}) on its parameters is not an easy task. In this section,
we present a feasible algorithm obtained by combining the EM
algorithm for CWM with that (with trimming and constraints)
introduced in \cite{GarG:ADAC13} \citep[see, also,][]{FriG13}:

\begin{enumerate}
  \item[1.] \textit{Initialization:}
  The algorithm is initialized several times by selecting
  different initial $\btheta^{(0)}=(\pi_1^{(0)},...,\pi_G^{(0)},\bmu_1^{(0)},...,\bmu_G^{(0)},
  \bSigma_1^{(0)},...,\bSigma_G^{(0)},b_1^{0(0)},...,b_G^{0(0)}, \bb_1^{(0)},..., \allowbreak\bb_G^{(0)}, \sigma_1^{2(0)}, ..., \sigma_G^{2(0)})$. 
  After drawing $d+2$ distinct observations for each group, we compute their
  sample means and sample covariance matrices as initial values for $\bmu_g^{(0)}$ and $\bSigma_g^{(0)}$. Additionally, $G$ ordinary least square regressions are
  carried out  to obtain initial $b_g^{0(0)}$ and $\bb_g^{(0)}$ regression parameters (G-inverse matrices
  are used if needed). The mean square errors of the $G$
  regressions are used to determine the initial $\sigma_g^{2(0)}$
  values.  If  $\bSigma_g^{(0)}$ and/or
  $\sigma_g^{2(0)}$ do not satisfy the required constraints
  (\ref{c_X}) and (\ref{c_eps}) then the procedure that will be described in
  Step 2.2 is applied to enforce them. Finally,
  weights $\pi_1^{(0)},...,\pi_G^{(0)}$ in the interval $(0,1)$ and summing up to 1 are
  randomly chosen.

  \item[2.] \textit{Trimmed EM steps:} Starting from each random initialization $\btheta^{(0)}$, the
following steps are alternatively executed until convergence or
until a maximum number of iterations is reached. The implementation
of trimming is clearly related to how ``concentration" steps
(C-steps) are carried out to implement high-breakdown robust methods
\citep[see, e.g.,][]{RouD98}.

  \begin{enumerate}
    \item[2.1.] \textit{E- and C-steps:}
    Let $\btheta^{(l)}$ be the parameters
   at iteration $l$, we compute
   $D_i=D(\bx_i,y_i;\btheta^{(l)})$ for $i=1,...,n$.
    After sorting these values, the notation
    $
    D_{(1)} \leq.... \leq
    D_{(n)}
    $
    is adopted. Let us consider the subset of indices $I \subset\{1,2,...,n\}$ defined
    as $
      I=\big\{i : D_{(i)}\geq D_{([n \alpha])}
      \big\} \label{eq:D(i)}.
 $
To update the parameters, we will take into account only the
observations with indices in $I$, by setting
 $ \tau_{ig}^{(l)} =D_{g}(\bx_i,y_i;\btheta^{(l)})/D(\bx_i,y_i;\btheta^{(l)})
$ for $i\in I$ and $\tau_{ig}^{(l)} =0$ for $i\notin I$. Note that
$\tau_{ig}^{(l)}$, for the observations with indices in $I$, are the
usual ``posterior probabilities" in the standard EM algorithm.

    \item[2.2.] \textit{M-step:} From these $\tau_{ig}$ values, we update the weight and mean parameters as
    $$
    \pi_g^{(l+1)}= \sum_{i=1}^n \tau_{ig}^{(l)}/[n(1-\alpha)]
    \quad \text{ and } \quad
    \bmu_g^{(l+1)}= \sum_{i=1}^n \tau_{ig}^{(l)}\bx_i \bigg/ \sum_{i=1}^n
    \tau_{ig}^{(l)}.
    $$
    The other parameters (regression and scatter ones) are initially updated by
    \begin{align*}
    T_g  & = \sum_{i=1}^n \tau_{ig}^{(l)}(\bx_i-\bmu_g^{(l+1)})(\bx_i-\bmu_g^{(l+1)})' \bigg/ \sum_{i=1}^n   \tau_{ig}^{(l)}, \\
      \bb_g^{(l+1)} &= \left(  \sum_{i=1}^n \tau_{ig}^{(l)}\bx_i\bx_i' \bigg/ \sum_{i=1}^n   \tau_{ig}^{(l)}      -      \left(    \sum_{i=1}^n \tau_{ig}^{(l)}\bx_i' \bigg/ \sum_{i=1}^n   \tau_{ig}^{(l)}     \right)^2          \right)^{-1}       \times \\ \nonumber
       & \left(    \sum_{i=1}^n \tau_{ig}^{(l)}y_i\bx_i' \bigg/ \sum_{i=1}^n   \tau_{ig}^{(l)}  -    \sum_{i=1}^n \tau_{ig}^{(l)}y_i \bigg/ \sum_{i=1}^n   \tau_{ig}^{(l)} \cdot \sum_{i=1}^n \tau_{ig}^{(l)}\bx_i' \bigg/ \sum_{i=1}^n   \tau_{ig}^{(l)}    \right),\nonumber \\
      b_{g}^{0(l+1)} & = \sum_{i=1}^n \tau_{ig}^{(l)}y_i \bigg/ \sum_{i=1}^n   \tau_{ig}^{(l)} - (\bb_g^{(l+1)})' \sum_{i=1}^n \tau_{ig}^{(l)}\bx_i' \bigg/ \sum_{i=1}^n   \tau_{ig}^{(l)} \\
   s_g^2 & =  \sum_{i=1}^n \tau_{ig}^{(l)}\bigg(y_i-(\bb_g^{(l+1)})' \bx_i-b_{g}^{0(l+1)}\bigg)^2 \bigg/ \sum_{i=1}^n   \tau_{ig}^{(l)}.
    \end{align*}

    Along the iterations, due to the updates,
    it may happen that the $T_g$ matrices and the $s_g^2$ values do not satisfy the required constraints for the scatter parameters.

To perform a constrained maximization of the sample covariance matrices, the singular-value decomposition of  $T_g=U_g'E_g  U_g$ is considered,
    with $U_g$ being an orthogonal matrix and
    $E_g=\text{diag}(e_{g1},e_{g2},\allowbreak..., e_{gd})$ a diagonal matrix. After defining the truncated eigenvalues as $
    [e_{gl}]_m^{X} = \min\big(c_X \cdot m, \max(e_{gl},m)\big),$
   with $m$ being some threshold value, then the scatter matrices are finally updated as
   $\bSigma_g^{(l+1)}=U_g'E_g^*U_g,$
   with $E_g^*=\text{diag}\left([e_{g1}]_{m_{\text{opt}}^X}^{X},[e_{g2}]_{m_{\text{opt}}^X}^{X},...,[e_{gp}]_{m_{\text{opt}}^X}^{X} \right)$
   and $m_{\text{opt}}^X$ minimizing the real valued function
   \begin{equation} \label{m1}
   m \mapsto \sum\limits_{g=1}^{G} \pi_g^{(l+1)} \sum\limits_{l=1}^{d}\left( \log
   \left( [e_{gl}]_m^{X}\right) +\frac{ e_{gl}}{[e_{gl}]_m^{X}}\right) .
   \end{equation}

   Analogously, in case that the $s_j^2$ parameters do not satisfy the constraint (\ref{c_eps}),
   we consider the truncated variances$
    [s_g^2]_m^{\varepsilon} = \min\big(c_{\varepsilon} \cdot m, \max(s_g^2,m)\big).$
    The variances of the error terms are finally updated as
   $\sigma_g^{2(l+1)}= [s_g^2]_{ m_{\text{opt}}^{\varepsilon} }^{\varepsilon},$
   with $m_{\text{opt}}^{\varepsilon}$ minimizing the real valued function
   \begin{equation} \label{m2}
   m \mapsto \sum\limits_{g=1}^{G} \pi_g^{(l+1)} \left( \log
   \left( [s_g^2]_m^{\varepsilon}\right) +\frac{ s_g^2}{[s_g^2]_m^{\varepsilon}}\right) .
   \end{equation}
Proposition 3.2 in \cite{FriG13} shows that $m_{\text{opt}}^X$ and
$m_{\text{opt}}^{\varepsilon}$ can be obtained, respectively, by
evaluating $2dG+1$ times the real valued function in (\ref{m1}) and
$2G+1$ times the real valued function in  (\ref{m2}).

  \end{enumerate}
  \item[3.] \textit{Choosing the best obtained solution:}
When the stopping criterium has been met, the value of the target
function (\ref{d1}) is computed. The parameters yielding the highest
value of the target function are returned as the final output of the
algorithm.
\end{enumerate}

\section{Constraints and trimming}\label{se4}

\subsection{Effect of  constraints}\label{se4_1}
The parameter $c_X$ controls the differences among scatters for the
normal distributions used as mixture components when modeling the
vector of covariates $\bX$. It also controls  the
deviations from sphericity in the multivariate case ($d>1$). As $c_X
< \infty$, we are avoiding that $|\bSigma_g|$ becomes arbitrarily
small, assuring a bounded contribution of $\phi_d(\bx_i
;\bmu_g,\bSigma_g)$ to the log-likelihood function in (\ref{d1}).
Moreover, a moderate value of $c_X$ avoids the detection of spurious
solutions, like in the case exemplified in Figure
\ref{fig:simdata2}. If we set  $c_X=1$,  then we force the
covariance matrices to satisfy the relation
$\bSigma_1=...=\bSigma_G=aI_d$ with $a>0$ and $I_d$ being the
identity matrix in $\mathbb{R}^d$. On the other hand, the larger the
value of $c_X$, the larger the differences among covariance matrices
modeling the mixture components  of $\bX$ could be.

\begin{figure}[!h]
\centering
\includegraphics[width=11.5cm]{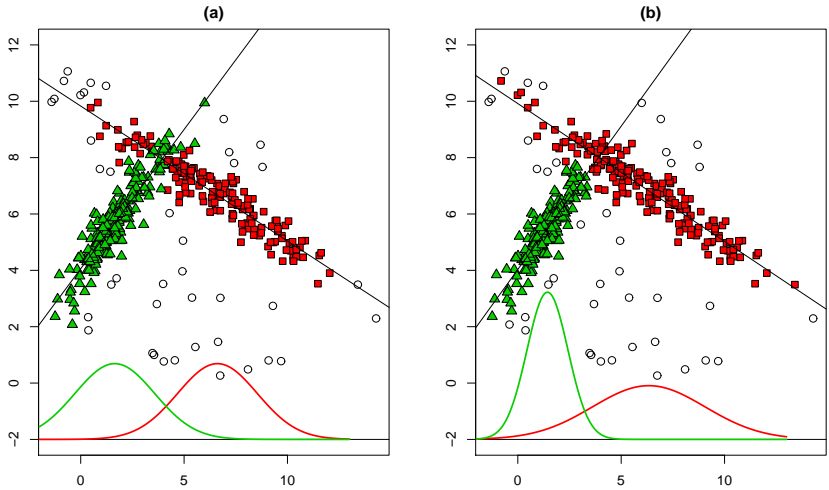}
\caption{\textit{Simdata4}: (a) Results for  $c_X=1$, that forces
equal scatters in the marginal distribution (the plotted densities, in the lower part of the figure,
represent the normal fitted components); (b) Results for $c_X=20$,
that allows different scatters. In both cases, $\alpha=0.1$ and
$c_{\varepsilon}=20$ have been chosen.}\label{fig:simdata4}
\end{figure}

For instance, consider the simulated data {\em Simdata4} 
in Figure \ref{fig:simdata4}, which is modeled according to either
$c_X=1$ or $c_X=20$, see Figure \ref{fig:simdata4},(a) and (b)
respectively. Note that the component variances ($\bSigma_1$ and
$\bSigma_2$ are positive real values because $d=1$) are forced to be
equal, i.e.:
$\bSigma_1=\bSigma_2$ in (a),
while
$\max\{\bSigma_1/\bSigma_2,\bSigma_2/\bSigma_1\}\leq 20$ holds in (b). The
densities of the normal distributions considered in the fitted
mixture
to model the $\bX$ distribution are also represented
below, to
illustrate their variances.

Our recommendation is to take $c_X>1$ without selecting huge values
for it. A sensible choice, for instance, is  $c_X=20$, as it worked
fairly well in  most of the cases we observed in practice, if the
explanatory variables are in similar scales.

\begin{figure}[!h]
\centering
\includegraphics[width=11.5cm]{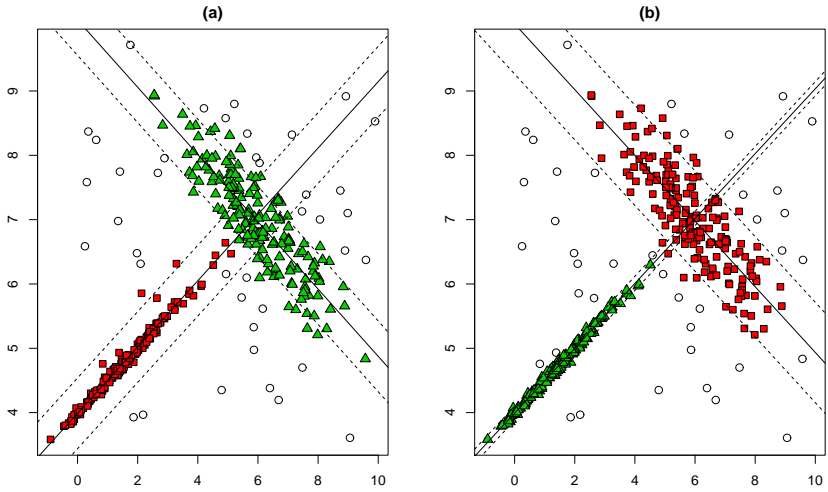}
\caption{\textit{Simdata5}: (a) Results for $c_{\varepsilon}=1$,
forcing equal variances in the error terms. (b) Results for a larger
$c_{\varepsilon}=20$ value. In both cases,  $\alpha=0.1$ and
$c_X=20$ have been chosen and bands of amplitude $\pm 2 \sigma_g$
 are shown.}\label{fig:simdata5}
\end{figure}

On the other hand, the constant $c_{\varepsilon}$ represents the
maximum ratio among the variances of the regression error terms.
Even if the ML estimation would be attracted by  solutions in which
some $\sigma_g^2 \rightarrow 0$, due to their high contribution by
means of $\phi(y_i ;\bb'_g \bx_i +b_{g}^0,\sigma_g^2)$ to the
maximization of the log-likelihood in (\ref{d1}), a choice of
$c_{\varepsilon}<\infty$ avoids that the algorithm fall into
singularities. Enforcing a  value $c_{\varepsilon}=1$ imposes the
strongest constraint $\sigma_1^2=...=\sigma_G^2$. For instance, let
us consider  {\em Simdata5} in Figure \ref{fig:simdata5}, which has
been generated from a CWM with $\sigma_1^2=0.5^2$ and
$\sigma_2^2=0.1^2$ ($\sigma_1^2/\sigma_2^2=25$). The results of
fitting the trimmed CWRM for this data set are also shown with
bands. Indeed, in specific applications, it is useful to take into
account such bands, centered at the fitted regression lines and with
amplitudes given by $\pm 2\sigma_g$,  i.e. twice the estimated
standard deviations of the regression error terms. A first solution
corresponding to $c_{\varepsilon}=1<25$ is given in Figure
\ref{fig:simdata5} (a),
while a second one corresponding to $c_{\varepsilon}=50>25$ is given
in panel \ref{fig:simdata5}(b). Notice the different amplitude  of
these bands. However, although different scatters can be effective
in many cases, a huge difference between them is not recommended, as
it can lead to  fit a few almost collinear observations.

An important feature of the proposed methodology is to provide a
different constraint for the eigenvalues of the matrices $\bSigma_g$
and for the variances of the error terms $\sigma_g^2$. This allows
to deal with  different scales in the explanatory and response
variables, which is common in many applications. On the other hand,
the procedure is not fully affine equivariant in the explanatory
variables, due to the considered constraints. However, if needed, it
is close to affine equivariance for large values of $c_X$.

It is well known, see e.g. \citet{Ingr:Mino:Vitt:Loca:2012}, that
the linear Gaussian CWM may be seen as included in the finite
mixture of Gaussian distributions when embedding it into a $d+1$
dimensional space. Also in the latter case, constraints are needed
to avoid singularities and to reduce the detection of spurious
solutions. However, constraints giving a completely symmetric
handling of the variability for the explanatory variables and for
the error terms are not always the best idea. For instance, as a way
to provide robustness, we could have considered the TCLUST
methodology \citep{GarG08} in the $d+1$ dimensional space which
needs the specification of a constant $c\geq 1$ to constraint the
maximal ratio among the $G\times(d+1)$ eigenvalues. Unfortunately,
Mixture of Regressions problems often require very high values for
the constant $c$ which do not always guarantee TCLUST to be
correctly protected against spurious solutions.

\begin{figure}[!h]
\centering
\includegraphics[width=7.5cm]{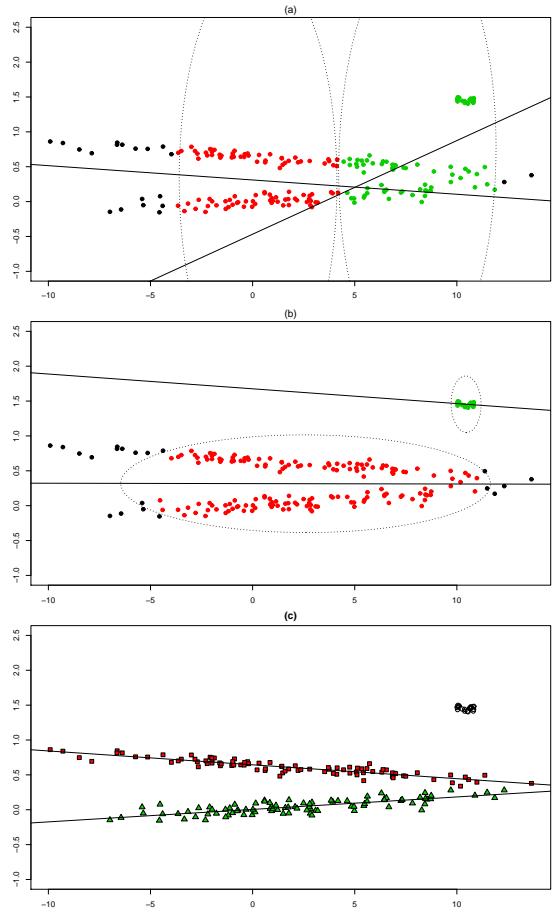}
\caption{\textit{Simdata6:} (a) TCLUST results with
$c=1.5$ and $\alpha=0.1$; (b) TCLUST results with
$c=500$ and $\alpha=0.1$; (c) Trimmed CWRM  fitting results with
$c_X=c_{\varepsilon}=1.5$ and $\alpha=0.1$.}\label{fig:simdata6}
\end{figure}

To illustrate the previous claims, let us consider {\em Simdata6}, of
size $n=200$, where 180 observations have been generated from a CWM
with two groups, and 20 observation have been included as
concentrated noise. The data set is plotted in Figure
\ref{fig:simdata6}, where panel (a) shows the results of applying
the TCLUST methodology with $c=1.5$ in dimension $d+1=2$. We can see
that the results are not satisfactory (the analogous of the
regression lines are the axes corresponding to the largest
eigenvalue of the $\bSigma_g$ matrices) and, therefore, higher $c$
values seem to be needed. But, higher $c$ values often yield the
detection of undesired spurious solutions. For instance, panel (b)
shows the results of applying TCLUST with $c=500$ with the detection
of a cluster only containing all noisy observations. On the other
hand, we can see that a proper fit is obtained in panel (c), when
applying the trimmed CWRM with $c_X=c_{\varepsilon}= 1.5$.

It is worthy to note that asymmetric constraints also underlies some
parameterizations already proposed in closely related problems as,
for instance, in \cite{DasR98} where the eigenvalues of the scatter
matrices corresponding to the $(d+1)$-dimensional fitted mixture
components are requested to be $\lambda_g\times
\{1,\alpha,...,\alpha\}$ with $\alpha <<1$.




\subsection{Effect of  trimming}\label{se4_2}
We start from the well-known Mixture of Regressions model and first
consider an easier trimming approach based on the maximization of
\begin{equation}\label{d2}
    \sum_{i=1}^n z(\bx_i, y_i)\log\left[\sum_{g=1}^G \phi(y_i ;\bb'_g \bx_i +b_{g}^0,\sigma_g^2)\pi_g \right],
\end{equation}
with $\sum_{i=1}^n z(\bx_i, y_i)=[n(1-\alpha)]$ and imposing a
constraint on the variances of the error terms
$\sigma_{g_1}^2/\sigma_{g_2}^2 \leq c_{\varepsilon}$ for $1\leq g_1,
g_2\leq G$. Notice that, in this case, the distribution of $\bX$ is
not taken into account, hence no trimming related to the $\bX$ model
is considered. This straightforward robust extension will be
referred to as trimmed Mixture of Regressions \citep{NeyF07,
GarS10}. Apart from the constraints, this approach reduces to the
traditional Mixture of Regressions when $\alpha=0$, and leads back
to the widely-applied Least Trimmed Squares (LTS) method \citep[see,
e.g.,][]{RouL87} when $G=1$ and $\alpha>0$. It protects against
large values of $(y_i-\bb'_g \bx_i -b_{g}^0)^2$, hence it is useful
to cope with many cases of data contamination which cause the
parameters $\bb_g$ ``breakdown", in absence of trimming. However, it
does not prevent the model estimation from the effects of ``bad"
leverage points, due to outliers in $\bx$. As it happens in ordinary
least squares regression, a few bad leverage points could provoke
very disappointing results.

\begin{figure}[!h]
\centering
\includegraphics[width=12cm]{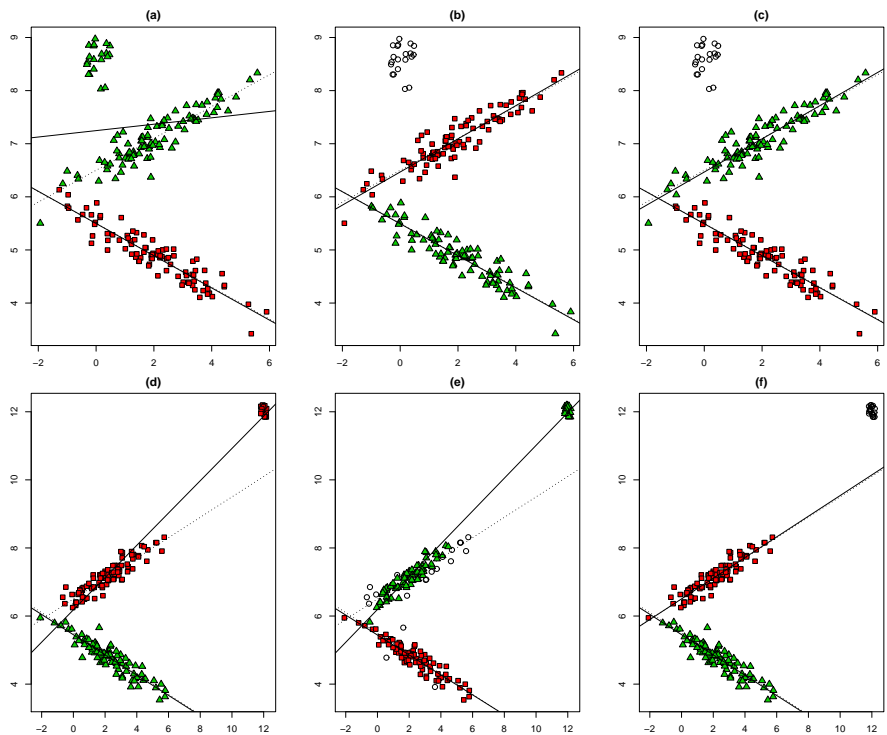}
\caption{{\em Simdata7} in the upper panels (a)-(c) and {\em
Simdata8} in the lower panels (d)-(f). (a) and (d) fitting the
(untrimmed) CWM; (b) and (e) fitting trimmed Mixture of Regressions;
(c) and (f) applying trimmed CWRM including a $10\%$ of
contamination. In particular, $\alpha=0.1$ and $c_{\varepsilon}=20$
are used in (b), (c), (e) and (f), while $c_X=20$ is used in (c) and
(f).}\label{fig:simdata7+8}
\end{figure}

For instance, consider the simulated datasets {\em Simdata7} and
{\em Simdata8} in Figure \ref{fig:simdata7+8}. Both datasets are
made of 180 observations drawn from a CWM with two groups and with
20 noisy observations generated by two different contamination
mechanisms. The leftmost panels in Figure \ref{fig:simdata7+8},(a)
and (d) show the results of fitting the standard CWM; the central
panels (b) and (e) concern trimmed Mixture of Regressions
($\alpha=0.1$) and, finally, the rightmost panels (c) and (f)
illustrate the proposed trimmed CWRM ($\alpha=0.1$). We can see that
the fit of the standard (untrimmed) CWM is strongly affected by the
contamination. Trimmed Mixtures of Regression are able to resist the
type of contamination in (b) but cannot afford outliers acting as
bad leverage points, as in (e). On the other hand, the use of
trimmed CWRM, as shown in (c) and (f), resists both types of
contamination. To avoid an unfair comparison, we have not included
remarkable differences in the $\bX$ distributions for the two main
groups (i.e., prior to contamination), but we can see in Figure
\ref{fig:simdata1} how the trimmed CWRM is able to deal with
components having different marginal distributions.

The problem of leverage points has been addressed in Robust
Regression by down-weighting influential observations as, for
instance, GM-estimators do \citep{KraW92}. In the context of
clusterwise regression, 
\cite{GarS10} proposed a ``second trimming", by fixing two trimming
parameters $\alpha_1$ and $\alpha_2$. Parameter $\alpha_1$ controls
the effect of outliers corresponding to large values of $(y_i-\bb'_g
\bx_i -b_{g}^0)^2$  while $\alpha_2$ aims at controlling leverage
points corresponding to outlying values on $\bx$. However, the
distinction between these two types of outliers is not always so
clear. On the other hand, the unified handling of outliers provided
by the trimmed CWRM simultaneously deals with both types of
outliers. As the probability to belong to a cluster is not a fixed
value, $\pi_g$, but depends also on the CWM weight
$\phi_d(x_i,\bmu_g,\bSigma_g)\pi_g$, trimming acts before on points
that lay on the farer contours of equiprobability (i.e.  sets of
points where the p.d.f. of the mixture takes a constant value) from
the cluster means. We are assuming that outliers are the points
$(\bx_i,y_i)$ with lower values of $D(\bx_i,y_i;\btheta)$, rather
than points with greater vertical distances $(y_i-\bb'_g \bx_i
-b_{g}^0)^2$.

Other alternatives to guard CWM against contamination are based on
the consideration of $t$-distributions, instead of normal ones, see
\cite{Ingr:Mino:Vitt:Loca:2012}. They provide a clear robustness
gain with respect to the Gaussian CWM. However, without trimming,
one single observation placed in a very remote position can still be
very harmful. In fact, we can make some components of $\bb_g$ to be
arbitrarily large or small, just by moving one single observation. A
small positive fraction of pointwise contamination can be very
dangerous too, even when it is not distant from the data. On the
other hand, the trimmed CWRM is more resistant to extreme
contaminations, because it does not make any assumption about how
outliers have been generated. Therefore, rather structured sources
of outliers  (and clearly not generated from a $t$-distribution) can
be handled, too.

Several methods can be also found in the literature aimed at
robustifying the Mixtures of Regressions model. Apart from those
based on trimming that have been previously cited, methods based on
M-estimation have been proposed in \cite{BaiY12} and extending
S-estimation in \cite{bashir2012robust}. \cite{song2013robust}
propose to model the  error terms by a
 Laplace distribution, while \cite{yao2014robust} suggest to employ the $t$ distribution. Although all these methods
 improve the robustness of the model,
 they do not
 model the marginal $\bX$
distribution. Therefore, they do not take advantage of this
information to detect the different mixture components and hence are
not able to cope with outliers both on $\bx$ and on $y$, acting as
bad leverage points. To overcome this issue, \cite{yao2014robust}
have recently proposed applying their robust Mixture of Regression
after using a trimming procedure (with high breakdown point) which
removes clear outliers on $\bx$. This initial trimming is
unfortunately done without considering the $Y$ variable, nor the
joint distribution in $(\bX,Y)$, corresponding to the different
mixture components. The MCD estimator, considered for this initial
trimming, is aimed at working on a single contaminated population
and can be troublesome for detecting outliers when the data set
includes different subpopulations.

In most of the applications, the true contamination level is
unknown. Therefore, it makes sense to consider a preventive (higher
than needed) trimming level $\alpha$. 
This could lead to wrongly trimmed observations, but the ``cores" of
the clusters and sensible approximations of the regression lines are
most of the times correctly found. Starting from them, it is not
difficult to recover wrongly trimmed observations, by resorting to
Mahalanobis distances and diagnostic regression tools \citep[see
Section 7 in][]{GarS10}.

\section{Real data examples}\label{se5}

\subsection{Tone data}\label{se5_1}

\begin{figure}[!ht]
\centering
\includegraphics[width=10.5cm]{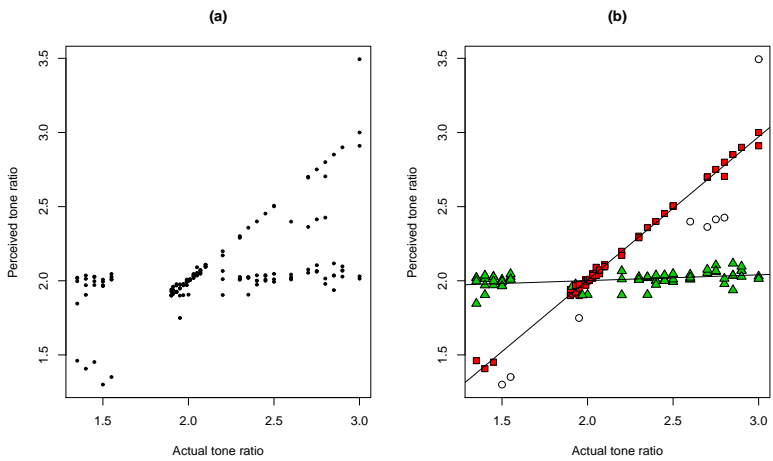}
\caption{{\em Tone data:} (a) Data set; (b) Trimmed CWRM fitting
with $\alpha=0.05$ and $c_X=c_{\varepsilon}=20$.}
\label{fig:tonedata}
\end{figure}

This data set comes from an experiment in music perception
introduced in \cite{CoeE84} which has been analyzed in many papers
concerning Mixtures of Regression, \citep[see, e.g.][]{deVR89} and
their robust versions \citep{{schlittgen2011weighted},
{hennig2002fixed}, {BaiY12}, {bashir2012robust}, {song2013robust},
{yao2014robust}}. This data set is shown in Figure
\ref{fig:tonedata}(a) and the result of applying the trimmed CWRM in
(b). We can see that the two main groups (interval memory judgement
and partial matching) can be detected by applying the trimmed CWRM.
Furthermore, $\alpha=0.05$ allows to detect a fraction of outlying
observations, within the partial matching group, exhibiting a clear
different behavior.

The type of outliers included in this data set are not very harmful
and, thus, no dramatic differences can be expected in terms of the
estimated parameters, when using any (robust) Mixture of Regressions
approach. So, we will proceed to artificially contaminate the data
and use it as a benchmark for the effects of leverage points added
through pointwise contamination. This has been already done by
\citet{BaiY12}, who introduced a $6\%$ of contamination at $(0,4)$,
when applying an M-estimation approach. In our case, we will use a
more complete contamination scheme by adding $9\%$ of point
contamination, placed around points $(2.5,5)$, $(6,4)$, $(0,0.5)$
and $(5,2.5)$, successively. The first location, $(2.5,5)$ is a
regression outlier, while the remaining three are leverage points.

\begin{table} [h]
\begin{tabular}{cclcclc}
\hline
 Contamination & & Trimmed CWRM   & Discarded  & & Trimmed MR &  Discarded   \\
     location& &  constants  &  outliers   & &  constants &  outliers    \\
    \hline
   $(2.5,5)$ & & $c_X=c_{\varepsilon}=1$  & Yes  & & $c_{\varepsilon}=1$  & Yes  \\
    & &  $c_X=c_{\varepsilon}=10^3$ & No  & &  $c_{\varepsilon}=10^3$ &  Yes  \\
    & &  $c_X=c_{\varepsilon}=10^{10}$ & No  & & $c_{\varepsilon}=10^{10}$  &  No \\
   $(6,4)$ & & $c_X=c_{\varepsilon}=1$  & Yes  & & $c_{\varepsilon}=1$  & No  \\
    & & $c_X=c_{\varepsilon}=10^3$  & No  & &  $c_{\varepsilon}=10^3$  & No \\
    & & $c_X=c_{\varepsilon}=10^{10}$  & No   & & $c_{\varepsilon}=10^{10}$  & No  \\
   $(0,0.5)$ & & $c_X=c_{\varepsilon}=1$  & Yes  & & $c_{\varepsilon}=1$  &  No \\
    & & $c_X=c_{\varepsilon}=10^3$  & Yes  & & $c_{\varepsilon}=10^3$  & No  \\
    & & $c_X=c_{\varepsilon}=10^{10}$  & No  & &  $c_{\varepsilon}=10^{10}$ & No \\
   $(5,2.5)$ & & $c_X=c_{\varepsilon}=1$  & Yes  & &  $c_{\varepsilon}=1$ & No  \\
    & & $c_X=c_{\varepsilon}=10^3$  & No  & &  $c_{\varepsilon}=10^3$ & No  \\
    & & $c_X=c_{\varepsilon}=10^{10}$  & No  & & $c_{\varepsilon}=10^{10}$  & No
    \\
    \hline
\end{tabular}
\caption{{\em Tone data:}  Performance comparison between the
trimmed CWRM methodology and trimmed Mixture of Regressions (trimmed
MR) with an $\alpha=0.1$ trimming level.}\label{table1}
\end{table}

Table \ref{table1} summarizes the performance of the proposed
trimmed CWRM  and the trimmed Mixture of Regressions (trimmed MR)
presented in Section \ref{se4_2}, both with an $\alpha=0.1$ trimming
level, for different values of the constraints factors $c_X$ and
$c_{\varepsilon}$, and labeling  by ``Yes"/``No" the cases in which
the trimming level allows/does not allow to discard all the noisy
observations. We can see that only the use of the trimmed CWRM with
$\alpha=0.1$ and with both constants fixed at their most restrictive
values is able to cope with the contamination in all the considered
scenarios.

\subsection{Students' heights and weights}\label{se5_2}

The data set in this example is based on students answers to a
questionnaire including simple questions about anthropometric
measurements. Due to the way in which the dataset has been
collected, it contains outliers, as some students did not seriously
answer the questions, or gave bad interpretations of the measurement
units, etc. Here, we focus on the relationship between two variables
in the data set, namely ``Height" ($X$) in cm and ``Weight" ($Y$) in
Kg. Although gender was also considered in the study, we will ignore
it, to test the ability of our methodology to classify the
individuals and to estimate the two underlying regression models,
one for each gender, in presence of an important amount of severe
outliers.

\begin{figure}[!h]
\centering
\includegraphics[width=11.5cm,height=14.5cm]{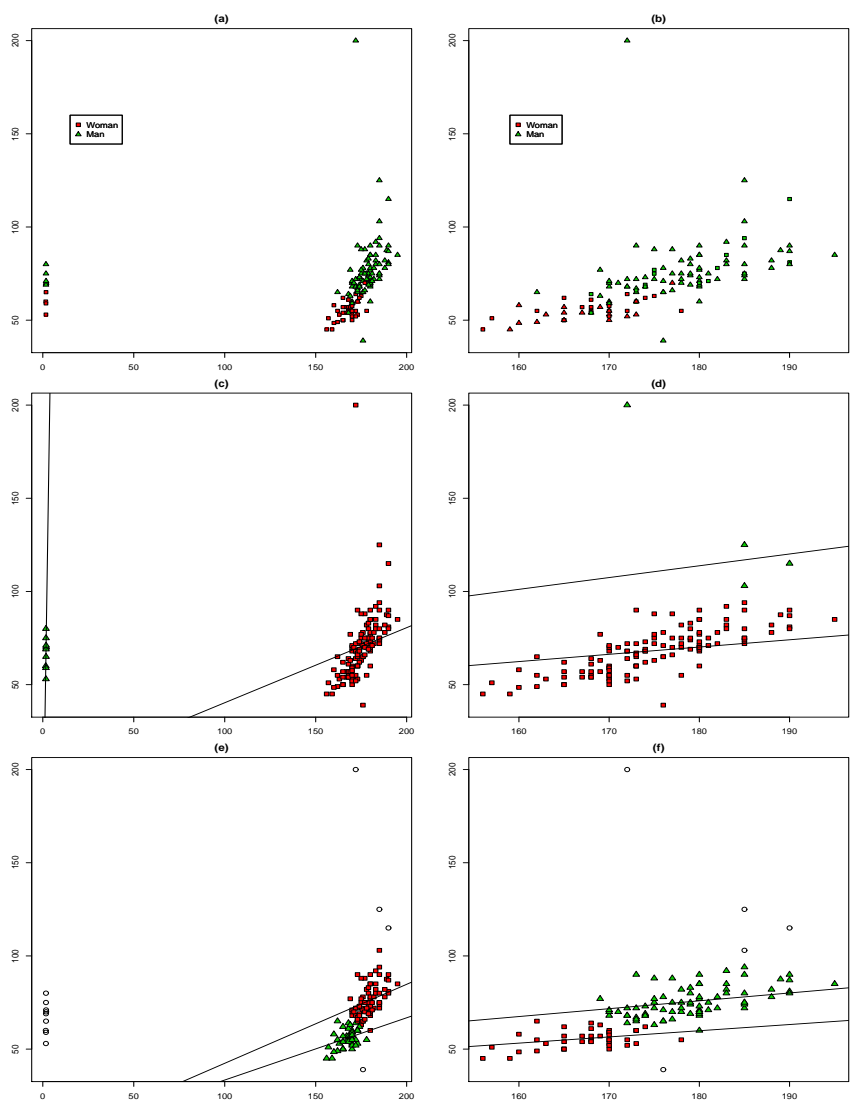}
\caption{{\em Student data: (a) ``Students' heights and weights"
data. (b) Cleaned data set obtained by deleting the outliers due to
wrong measurement scale for  ``height". Effects of trimming and
restrictions on CRWM results: (c) untrimmed and almost unrestricted:
$\alpha=0$ and $c_X=c_{\varepsilon}=10^{10}$; (d) untrimmed and
almost unrestricted: $\alpha=0$ and $c_X=c_{\varepsilon}=10^{10}$
for the cleaned data set; (e) trimmed and constrained: $\alpha=0.1$
and $c_X=c_{\varepsilon}=20$; (f)
 trimmed and constrained: $\alpha=0.04$ and $c_X=c_{\varepsilon}=20$ for the cleaned data
set}}\label{fig:student+CRWM}
\end{figure}

Figure \ref{fig:student+CRWM}(a) shows the original data set (which
will be referred to as {\em Student data}) with the true gender
assignments, while in (b)  we have eliminated the points
corresponding to a wrong scale in height (students reporting height
in meters instead of centimeters), to emphasize the different linear
patterns. Several implausible weight values can be also seen. Figure
\ref{fig:student+CRWM}(c) shows the results corresponding to the fit
of the CWM (when $\alpha=0$ and $c_x=c_\epsilon=10^{10}$, i.e., no
trimming and almost unrestricted). We can see that one of the
regression lines is capturing the artificial group, almost
collinear, having anomalous height values. Consequently, the main
groups are joined together and the classification error rate is very
high. On the other hand, Figure \ref{fig:student+CRWM}(e) shows the
result of applying the trimmed CWRM with $\alpha=0.1$ and moderate
values of the constraints. Restrictions now avoid that the method
falls into the previously obtained spurious solution, generated by
the almost collinear outliers (wrong measurement units) and these
points are trimmed off, together with other data points exhibiting
atypical weight values. The classification error rate for untrimmed
observations is just $12\%$. Figures \ref{fig:student+CRWM}(d) and
(f) show the data set after eliminating the points with
wrong units for the height. In Figure \ref{fig:student+CRWM}(d), we
can see that using the CWM, even in this cleaned data set, again
fails to detect the true groups. On the contrary, we can see in (f)
that the trimmed CWRM with $\alpha=0.04$ and moderate values of
$c_X$ and $c_{\epsilon}$ provides sensible results. It is true that
simple visual inspection could have served to ``clean" this data set
but this is surely not the case when dealing with more complex/high
dimensional data sets on when carrying out fully unsupervised data
analyses.

\section{Concluding remarks}\label{se6}
The present work is centered on the wide family of Gaussian CWMs,
that   received a growing attention in the recent literature.
However, like it happens for many other models which depend on normal assumptions,
the ML estimation for CWM suffers from a lack of robustness. Moreover, the
problem statement in terms of the likelihood maximization  is not
well-posed, without constraints. Hence, here we have presented a new
estimation framework for the linear Gaussian CWM based on trimming
and constraints, to achieve robustness, identify and discard
outliers, circumvent the likelihood singularities and reduce the
detection of spurious solutions.

Numerical studies, based on both simulated and real data, show that
the new proposal drives the estimation procedure to discard even
strongly concentrated contaminating observations, acting as bad
leverage points, which are so harmful in the framework of Mixtures
of Regressions. Apart from the effectiveness of the proposed methodology to resist to any kind of outliers, we have also shown that  a theoretically well defined
mathematical and statistical problem underlies it. The existence of
optima for both the population and the sample problem have been
established, and the consistency of the sample solution to the
population one has been provided.

Further research could be focused on tuning the choice of the involved
parameters. This is a complex task, as these parameters are
clearly interrelated.
 For instance, a high trimming level $\alpha$
could lead to smaller $G$ values, since components with fewer
observations may be trimmed off. Moreover, larger values of $c_X$
and $c_{\varepsilon}$ could lead to higher values of $G$, since more
components with few observations, but close to collinearity, may be
detected. Our suggestion is that the researcher must provide in
advance part of these parameters (as a way of specifying the type of
clusters expected from the data) and, then, some data-dependent
diagnostic can be used to make appropriate choices for the rest of
parameters. 
The use of trimmed BIC notions \citep{NeyF07} or the adaptation of
some graphical tools, as in \cite{GarG11}, can be useful for this
purpose.

\begin{acknowledgements}
This research is partially supported by the Spanish Ministerio de Ciencia e Innovaci—n, grant MTM2011-28657-C02-01, by Consejer'a de Educaci—n de la Junta de Castilla y Le—n, grant VA212U13, and by grant FAR 2013 from the University of Milano-Bicocca.
\end{acknowledgements}

\section*{Appendix}

The following section is organized into four parts: part A contains technical lemmas useful for the proof of the existence of the maximizer $\btheta$ for $L(\btheta,P)$ (Proposition 3.2.1)  which is established in part B; part C shows preliminary results needed to show the consistency of $\hat{\btheta}$ as an estimator for $\btheta$ (Proposition 3.2.2), which is then proved in part D.

\subsection*{\bf{Part A: Preliminary results in view of Proposition 3.2.1}}

Four technical lemmas will be needed before attacking the proof of
Proposition 3.2.1.

First of all, let us remark that, given the definition of
$L(\btheta,P)$, there exist
 sequences $\{\btheta_n\}_{n=1}^{\infty}$ with
\begin{equation}\label{a0}
    \btheta_n=(\pi_1^n,...,\pi_G^n,\allowbreak \bmu_1^n,...,\bmu_G^n,
  \bSigma_1^n,...,\bSigma_G^n,b_1^{0,n},...,b_G^{0,n}, \bb_1^n,..., \bb_G^n,\allowbreak \sigma_1^{2,n}, ...,
  \sigma_G^{2,n}),
\end{equation}
  and $\btheta_n \in \Theta_{c_X,c_{\varepsilon}}$
and such that
  \begin{equation}\label{a1}
    \lim_{n\rightarrow \infty}L(\btheta_n,P)=\sup_{\btheta\in
    \Theta_{c_X,c_{\varepsilon}}} L(\btheta,P)>-\infty
  \end{equation}
 (the boundedness from below is obtained just by considering the set $A$
 as being a ball centered at $(\mathbf{0},0)$ with
 $P[A] \geq 1-\alpha$,
 $\pi_1=1$, $\bmu_1=\mathbf{0}$, $\bSigma_1=I_d$, $b_1^0=0$ and $\bb_1=\mathbf{0}$).

  The proof of the existence
  will be done by proving that we can
 obtain a convergent subsequence
 extracted from
  $\{\btheta_n\}_{n=1}^{\infty}$ satisfying (\ref{a1}),
 and whose limit  $\btheta_0$ is optimal for  $P$.

Let us begin with Lemma \ref{l1}, which provides  a uniformly
bounded representation of the regression coefficients,  even in case
of local collinearity, without loosing their properties in the
evaluation of the target function.

\begin{lemma}
\label{l1} Let $\{b_n^0\}_{n=1}^{\infty}$ be a sequence in
$\mathbb{R}$, $\{\bb_n\}_{n=1}^{\infty}$ be a sequence in
$\mathbb{R}^d$ and $\{A_n\}_{n=1}^{\infty}$ be a sequence of sets in
$\mathbb{R}^{d+1}$ verifying
\begin{equation}\label{le1_1}
    \lim\sup_n P[A_n]>0
\end{equation}
and such that
\begin{equation}\label{le1_2}
    \lim\sup_n E_P \big[ |b_n^0+\bb_n' \bX - Y|^2 I_{A_n}(\bX,Y) \big]
    <\infty.
\end{equation}
Then, we can extract subsequences  $\{b_{n_k}^0\}_{k=1}^{\infty}$,
$\{\bb_{n_k}\}_{k=1}^{\infty}$ and $\{A_{n_k}\}_{k=1}^{\infty}$ from
them and define new sequences $\{d_k^0\}_{k=1}^{\infty}$,
$\{\bd_k\}_{k=1}^{\infty}$ and $\{D_k\}_{k=1}^{\infty}$  which
satisfy $D_k \subseteq A_{n_k}$, $P[A_{n_k} \setminus
D_k]\rightarrow 0$, $d_{n_k}^0 \rightarrow d^0\in \mathbb{R}$,
$\bd_{n_k} \rightarrow \bd\in \mathbb{R}^d$ and such that
\begin{equation}\label{le1_3}
(b_{n_k}^0+\bb_{n_k}' \bX - Y) I_{D_k}(\bX,Y) = (d_k^0+\bd_k' \bX -
Y)I_{D_k}(\bX,Y),\text{ }P\text{-a.s.},
\end{equation}
for every $k\geq 1$.
\end{lemma}

\noindent{\sl Proof:} To simplify the proof, w.l.o.g., we will use
the same notation for the subsequences as that used for the original
sequences. If the sequences $\{b_n^0\}_{n=1}^{\infty}$ and
$\{\bb_n\}_{n=1}^{\infty}$ are bounded, then we just need to extract
convergent subsequences and
set $D_n=A_n$. So, let us assume that either one or both sequences
are unbounded, and consider a sequence of compact sets
$\{K_n\}_{n=1}^{\infty}$ such that $K_n\uparrow \mathbb{R}^{d+1}$.
Let $\{\bv_{n_l}\}_{l=1}^{d}$
be the normalized eigenvectors 
obtained from the spectral decomposition of the matrices
$\{\text{Var}_P[\bX/A_n\cap K_n]\}_{n=1}^{\infty}$  (we use
 $E_P[\cdot/A]$ and $\text{Var}_P[\cdot/A]$ for denoting
$E_P[\cdot/(\bX,Y)\in A]$ and $\text{Var}_P[\cdot/(\bX,Y)\in A]$).

Now, let us suppose that there exists a  direction $\bv_{n_l}$ such
that $\text{Var}_P[\bv_{n_l}'\bX/A_n\cap K_n]\rightarrow 0$ then
take $H$ with $0\leq H < d$ and such that
$\text{Var}_P[\bv_{n_l}'\bX/A_n\cap K_n]\rightarrow 0$ for every
$l\geq H+1$, after a possible reordering of the coordinates. In this
case, there also exist points $\{\bu_{n_l}\}_{l=H+1}^{d}$ in
$\mathbb{R}^d$ and a sequence $\varepsilon_n\downarrow 0$ which must
satisfy $E_P[|\bv_{n_l}'(\bX-\bu_{n_l})|>\varepsilon_n/A_n\cap
K_n]\rightarrow 0$ for every $l\geq H+1$. The $\bv_{n_l}$ are
bounded (unitary vectors) and the $\bu_{n_l}$ must be bounded too
(because, otherwise, $\bX$ would not be tight). Therefore, there
exist subsequences, that will be denoted as the original ones, such
that $\bv_{n_l}\rightarrow \bv_l\in \mathbb{R}^d$,
$\bu_{n_l}\rightarrow\bu_l\in \mathbb{R}^d$ and
$P[|\bv_l'(\bX-\bu_l)|>0/A_n\cap K_n]\rightarrow 0$ for every $l\geq
H+1$.

Let us now define $D_n=A_n \cap K_n \cap_{l=H+1}^d
\{\bv_l'(\bX-\bu_l)=0\}$ which trivially verifies $D_n \subset A_n$
and that $P[A_n \setminus D_n]\rightarrow 0$. We can rewrite
$$
    b_n^0+\bb_n' \bx=
b_n^0+\sum_{l=1}^H\bb_n' \bv_l \bv_l' \bx+\sum_{l=H+1}^d\bb_n' \bv_l
\bv_l' \bx.
$$
and
set $d_n^0=b_n^0+\sum_{l=H+1}^d\bb_n' \bu_l $ and
$\bd_n=\sum_{l=1}^H \bb_n'\bv_l \bv_l'$ for $H>0$ (while we set
$\bd_n=\mathbf{0}$ when $H=0$). Then (\ref{le1_3}) trivially holds
and it can be shown that $\{d_n^0\}_{n=1}^{\infty}$ and
$\{\bd_n\}_{n=1}^{\infty}$ are bounded sequences. This follows from
the fact that (\ref{le1_2}) guarantees that $\{(b_n^0+\bb_n' \bX -
Y)I_{D_n}(\bX,Y)\}_{n=1}^{\infty}$ is a tight sequence. Notice that
we could see that the previous tightness property would be
contradicted if any of the $\{d_n^0\}_{n=1}^{\infty}$ and
$\{\bd_n\}_{n=1}^{\infty}$ were unbounded by seeing that
$\bZ=(Z_1,...,Z_H)$ with $Z_l=\bv_l' \bx$ satisfies
$\det(\text{Var}_P[\bZ/A_n\cap K_n])>0$ and 
$\bd_n'\bx=\sum_{l=1}^H \bb_n'\bv_l Z_l$.

Finally, whenever none of the sequences
$\text{Var}_P[\bv_{n_l}'\bX/A_n\cap K_n]$ converges to 0, we can
consider the representation $b_n^0+\bb_n' \bx=
b_n^0+\sum_{l=1}^H\bb_n' \bv_l \bv_l' \bx$ and  the result would be
proven in this case, too, following similar arguments as before.
$\Box$\smallskip

The following Lemma \ref{l2} assures that, under the usual
assumption on $P$, the associated fitted trimmed CWMs could not be
arbitrarily close to a degenerated model concentrated on $G$ points,
nor on $G$ regression hyperplanes.
\begin{lemma}
\label{l2}  Let $P$ be a distribution in $\mathbb{R}^{d+1}$
satisfiying \emph{(PR)}:
\begin{enumerate}
  \item[\emph{(a)}] For every $b_g^0 \in \mathbb{R}$,
  $\bb_g \in \mathbb{R}^d$ and
$A \subseteq \mathbb{R}^{d+1}$ with $P[A]=1-\alpha$, there exists
$\delta >0$ such that
$$
E_P\bigg[ \min_{g=1,...,G}|b_g^0+\bb_g' \bX - Y|^2 I_A(\bX,Y)\bigg]
\geq \delta.
$$

  \item[\emph{(b)}] For every set of $G$ points $\{\mathbf{\bmu}_1,...,\mathbf{\bmu}_G\}\subset \mathbb{R}^d$ and $A \subseteq \mathbb{R}^{d+1}$ with
  $P[A]=1-\alpha$,
there exists $\delta >0$ such that
$$
E_P\bigg[ \min_{g=1,...,G}\| \bX - \mathbf{\bmu}_g \|^2
I_A(\bX,Y)\bigg] \geq \delta.
$$
\end{enumerate}

\end{lemma}

\noindent{\sl Proof of (a):} Let us suppose that 
$\delta$ does not exist. Then, we can choose sequences
$\{A_n\}_{n=1}^{\infty}$, $\{b_g^{0,n}\}_{n=1}^{\infty}$ and
$\{\bb_g^n\}_{n=1}^{\infty}$ such that
\begin{equation}\label{a1b}
    E_P\bigg[ \min_{g=1,...,G}|b_g^{0,n}+(\bb_g^n)' \bx - y|^2
    I_{A_n}(\bx,y)\bigg]\rightarrow 0 \text{ with }P[A_n]\rightarrow 1-\alpha.
\end{equation}
Moreover, we can replace the sets $A_n$  in (\ref{a1b}), by the data
sets $$ A_n^*= \big\{(\bx,y):\min_{g=1,...,G} |b_g^{0,n}+(\bb_g^n)'
\bx - y|^2\leq \min\{r_{\alpha}^n,\varepsilon\}\big\},$$ where
$r_{\alpha}^n=\inf_u\{P[(\bx,y):\min_{g=1,...,G}\allowbreak
|b_g^{0,n}+(\bb_g^n)' \bx - y|^2\leq u]\geq 1- \alpha\}$ and we also
have the same convergence as in (\ref{a1b}), 
with $P[A_n^*] \rightarrow 1-\alpha$ for any fixed choice of
$\varepsilon>0$. Then, take
$$\hspace*{-2cm}A_g^n=\bigg\{(\bx,y)\in A_n^*: |b_g^{0,n}+(\bb_g^n)' \bx -
y|=\min_{j=1,...,G} |b_j^{0,n}+(\bb_j^n)' \bx - y|\bigg\},$$ and, we
can see that there exists at least one $g$ such that
$P[A_g^n]\rightarrow p_g
>0$ through a subsequence (because $P[A_n^*]=\sum_{g=1,...,G}
P[A_g^n]\rightarrow 1-\alpha$). Thus, consider a reordering of
$\{1,...,G\}$ such that $P[A_g^n]\rightarrow p_g >0$ for every
$g\in\{1,...,H\}$ (for an appropriate subsequence, if needed). If
$A_n^{**}=\cup_{g=1}^H A_g^n$, then
$$
E_P\bigg[ \min_{g=1,...,G}|b_g^{0,n}+(\bb_g^n)' \bX - Y|^2
    I_{A_n^{**}}(\bX,Y)\bigg]$$
    $$\quad\quad\quad\quad\quad\quad=\sum_{g=1}^H E_P\bigg[|b_g^{0,n}+(\bb_g^n)' \bX - Y|^2
    I_{A^n_g}(\bX,Y)\bigg]
$$
and $P[A_n^{**}]\rightarrow 1-\alpha$. For every $g\in\{1,...,H\}$,
the $A_g^n$, $b_g^{0,n}$ and $\bb_g^n$ satisfy the conditions needed
to apply Lemma \ref{l1} and, therefore, we can replace them by
$D_g^n$, $d_g^{0,n}$ and $\bd_g^n$ satisfying $D_g^n\subset A_g^n$,
$P[A_g^n\setminus D_g^n]\rightarrow 0$, $d_g^{0,n}\rightarrow
d_g^0\in \mathbb{R}$ and $\bd_g^n\rightarrow \bd_g^0\in
\mathbb{R}^d$ and (\ref{le1_3}).

Now, take $B_n=\cup_{g=1,...,H}D_g^n\cap
\{(\bx,y):\min_{g=1,...,G}|d_g^{0,n}+(\bd_g^n)' \bx - y|^2\leq
\varepsilon\}$ for a fixed $\varepsilon$, with $P[B_n]\rightarrow
1-\alpha$. We thus have the pointwise convergence
$$
  \min_{g=1,...,H}|d_g^{0,n}+(\bd_g^n)' \bx - y|^2
    I_{B_n}(\bx,y)\rightarrow  \min_{g=1,...,H}|d_g^{0}+(\bd_g^0)' \bx - y|^2
    I_{B_0}(\bx,y),
$$
for any $B_0\subset \mathbb{R}^{d+1}$ with $P[B_0]=1- \alpha$, and
the uniform bound $ \min_{g=1,...,H}|d_g^{0,n}+(\bd_g^n)' \bX - Y|^2
    I_{B_n}(\bx,y) \leq \varepsilon.
$ Then, the dominated convergence theorem implies
$$
E_p \bigg[\min_{g=1,...,H}|d_g^{0,n}+(\bd_g^n)' \bX - Y|^2
    I_{B_n}(\bX,Y)\bigg]$$
    $$\quad\quad\quad\quad\quad\quad \rightarrow E_p \bigg[\min_{g=1,...,H}|d_g^{0}+(\bd_g^0)' \bX - Y|^2
    I_{B_0}(\bX,Y)\bigg].
$$
The latter convergence and (\ref{a1b}) would prove that $$E_p
\bigg[\min_{g=1,...,H}|d_g^{0}+(\bd_g^0)' \bX - Y|^2
    I_{B_0}(\bX,Y)\bigg]=0,$$
    implying that the distribution
    $P$ is concentrated on $G$ regression hyperplanes after removing a
    proportion $\alpha$ of the probability mass and this would contradict (PR).\smallskip

\noindent{\sl Proof of (b):} The proof of this results mimics the
steps followed in the proof of (a). We start by assuming the
existence of subsequences $\{A_n\}_{n=1}^{\infty}$ and
$\{\bmu_g^n\}_{n=1}^{\infty}$ such that
$$
    E_P\bigg[ \min_{g=1,...,G}\|\bx - \bmu_g^n\|^2
    I_{A_n}(\bx,y)\bigg]\rightarrow 0 \text{ with }P[A_n]\rightarrow 1-\alpha.
$$
and we would end up by seeing that the support $\bX$ is concentrated
in $G$ points in $\mathbb{R}^d$. In fact, the proof is easier
because only the tightness of $P$ is needed (Lemma \ref{l1} is no
longer required, here). $\Box$\smallskip

 Now,
 since $[0,1]^G$ is a compact set, we can trivially choose a
 subsequence of $\{\btheta_n\}_{n=1}^{\infty}$ such that $
 \pi_g^n \rightarrow \pi_g \in [0,1]\text{ for }1\leq g \leq G.
 $ With respect to the scatter matrices and the variances of the error
 terms, we have the following possibilities:
\begin{eqnarray}\nonumber
   && \text{(S1) }  \bSigma_g^n \rightarrow \bSigma_g \text{ for }1\leq g \leq G \text{ with } \bSigma_g \text{ being p.s.d. matrices}\\\nonumber
   && \text{(S2) } \min_{g=1,...,G} \min_{l=1,...,d} \lambda_l(\bSigma_g^n) \rightarrow \infty  \\\nonumber
   && \text{(S3) } \max_{g=1,...,G} \max_{l=1,...,d} \lambda_l(\bSigma_g^n) \rightarrow 0  \\\nonumber
   && \text{(V1) } \sigma_g^{2,n} \rightarrow \sigma_g^2 \text{ for }1\leq g \leq G \text{ with } \sigma_g >0\\\nonumber
   && \text{(V2) } \min_{g=1,...,G} \sigma_g^{2,n} \rightarrow \infty \\\nonumber
   && \text{(V3) } \max_{g=1,...,G} \sigma_g^{2,n} \rightarrow 0
\end{eqnarray}
Given that $\btheta_n\in \Theta_{c_X,c_{\varepsilon}}$, only one of
the convergences in S1-S3 and only one in V1-V3 are possible, and
the following Lemma \ref{l3} will further delimitate to the bounded
results, based on constraints (5) and (6).
\begin{lemma}
\label{l3} If $\{\btheta_n\}_{n=1}^{\infty}\subset
\Theta_{c_X,c_{\varepsilon}}$ converges toward the supremum of
$L(\btheta,P)$, and  \emph{(PR)} holds for $P$, then only
convergences \emph{(S1)} and \emph{(V1)} are possible.

\end{lemma}

\noindent{\sl Proof:} We have that
 $L(\btheta_n;P)$ can be
 bounded from above by
$$
   - \frac{1}{2}\left[ \log\bigg( \min_g \sigma_g^{2,n} \bigg)P[A(\btheta_n)] +   \frac{E_P\big[\min_g|b_g^{0,n}+(\bb_g^n)'\bX-Y|^2 I_{A(\btheta_n)}(\bX,Y)\big]}{\max_g \sigma_g^{2,n}}   \right] \\
$$
$$ \hspace*{0.15cm} - \frac{1}{2}\left[ \log\bigg( \min_g \min_l \lambda_l(\bSigma_g^n)
\bigg)P[A(\btheta_n)]d +
\frac{E_P\big[\min_g\|\bX-\bmu_g^n\|^2I_{A(\btheta_n)}(\bX,Y)\big]}{\max_g
\max_l \lambda_l(\bSigma_g^n) }
  \right]+C,
$$
where $C$ is a constant value, not depending on $\btheta_n$.

Therefore, given that $\btheta_n\in \Theta_{c_X,c_{\varepsilon}}$,
we see that the possible convergence of $L(\btheta_n;P)$ would
clearly depend on those for the sequences
\begin{equation}\label{a2}
   \log\bigg( \frac{\sigma_n^2}{c_{\varepsilon}} \bigg)P[A(\btheta_n)]+E_P\bigg[\min_g \big|b_g^{0,n}+(\bb_g^n)' \bX-Y\big|^2 I_{A(\btheta_n)}(\bX,Y)\bigg]\frac{1}{\sigma_n^2}
\end{equation}
and
\begin{equation}\label{a3}
    \log\bigg( \frac{\lambda_n}{c_X} \bigg)P[A(\btheta_n)]d+E_P\bigg[\min_g \|\bX-\bmu_g^n\|^2
    I_{A(\btheta_n)}(\bX,Y)\bigg]\frac{1}{\lambda_n},
\end{equation}
where $\lambda_n=\max_{g=1,...,G} \max_{l=1,...,d}
\lambda_l(\bSigma_g^n)$ and $\sigma_n^2=\max_{g=1,...,G}
\sigma_g^{2,n}$.

On the other hand, Lemma \ref{l2} implies that a constant $\delta>0$
can be chosen such that $E_P\big[\min_g |b_g^{0,n}+(\bb_g^n)'
\bX-Y|^2 I_{A_n}(\bX,Y)\big]$ and $E_P\big[\min_g \|\bX-\bmu_g\|^2
I_{A_n}(\bX,Y)\big]$ in (\ref{a2}) and (\ref{a3}) are uniformly
bounded from below by $\delta$. Therefore, other convergences
different from (S1) or (V1) would imply that $\lim_{n\rightarrow
\infty}L(\btheta_n,P)=-\infty$ and this would contradict (\ref{a1}).
$\Box$ \smallskip

Lemma 4, stated below, shows that we can always find a subsequence
$\{\btheta_n\}_{n=1}^\infty$ with converging parameters for at least
one mixture component, with weight $\pi_g^n$ converging toward a
strictly positive value.
\begin{lemma}
\label{l4} There exists a sequence $\{\btheta_n\}_{n=1}^{\infty}$
converging toward the supremum of $L(\btheta,P)$ and there exists
$H$ with $1\leq H\leq G$ such that
$$ \bmu_g^n \rightarrow \bmu_g, \quad b_g^{0,n}\rightarrow b_g^0, \quad \bb_g^n \rightarrow \bb_g \quad \textrm{and}
\quad \pi_g^n \rightarrow \pi_g >0 \quad \textrm{for every}  \quad
g\leq H$$ and such that the corresponding
$\{A(\btheta_n)\}_{n=1}^{\infty}$ sets  are uniformly bounded.
\end{lemma}

\noindent{\sl Proof:} Let us start from any
$\{\btheta_n\}_{n=1}^{\infty}$ converging toward the supremum of
$L(\btheta,P)$, and take $A_n=A(\btheta_n)$ and $$ A_g^n=\big\{
(\bx,y)\in A_n : D_g(\bx,y;\btheta)=\max_{j=1,...,G}
D_j(\bx,y;\btheta)\big\}$$ for $1\leq g \leq G$. Since
$P[A_g^n]\in[0,1]$, there exists a subsequence, denoted as the
original one, such that each $P[A_g^n]$ converges for $1\leq g \leq
G$. Moreover, after a proper reordering in the components of
$\btheta_n$, there exists $H^*\geq 1$ such that $P[A_g^n]\rightarrow
p_g>0$ for $1\leq g \leq H^*$. Note that this $H^*$ does exist
because otherwise we would have 
$P[A_n]=\sum_{g=1}^GP[A_g^n]\rightarrow 0$.

We can also find a convergent subsequence of $\bmu_g^n$ for every
$g\leq H^*$. Otherwise, for every $\eta$ with $0<\eta<p_g$, we could
take a ball $B_g$ centered at $(\mathbf{0},0)$ with
$P[B_g]>1-p_g+\eta$ and such that there exists $n_0$ with $P[B_g\cap
A_g^n]>\eta/2$ when $n\geq n_0$. Consequently, we would have
$E_P\big[ \| \bX-\bmu_g^n \|^2 I_{A_g^n}\big]\geq E_P\big[ \|
\bX-\bmu_g^n \|^2 I_{B_g\cap A_g^n}\big] \rightarrow \infty$ which
contradicts (\ref{a1}). Note that the contributions of the other
terms to $L(\btheta_n,P)$ are controlled, because of Lemma \ref{l3}.

From (\ref{a1}), we have $\lim\sup_n E_P\big[ |b_g^{0,n}+(\bb_g^n)'
\bX-Y|^2 I_{A_g^n}(\bX,Y)\big]<\infty$. This, together with the fact
that $\lim\sup_nP[A_g^n]=p_g>0$ for $g \leq H^*$, allows us to apply
again Lemma \ref{l1} to replace the $\{b_g^{0,n}\}$, $\{\bb_g^n\}$
and $\{A_g^n\}$ sequences by appropriated convergent sequences
$\{d_g^{0,n}\}$, $\{\bd_g^n\}$ and $\{D_g^n\}$. These convergences
also trivially imply that $\pi_g^n \rightarrow \pi_g
>0$ for $g\leq H^*$.

Other $g$ values could also satisfy these convergences (through
subsequences and possible alternative representations). In this
case, we consider $H\geq H^*$ such that all the convergences in the
statement of this Lemma hold for $g\leq H.$

To see that the $\{A(\btheta_n)\}_{n=1}^{\infty}$ are uniformly
bounded, recall that $A(\btheta_n)=\{(\bx,y):D(\bx,y;\btheta_n)\geq
R(\btheta_n,P)\}$ and let us introduce
$$\widetilde{R}(\btheta_n,P)=\sup_u\bigg\{ P\bigg[\max_{1\leq g\leq H}
D_g(\bX,Y;\btheta_n)\geq u\bigg] \geq 1-\alpha \bigg\}.$$ Given that
$D(\bx,y;\btheta_n)\geq \max_g D_g(\bx,y;\btheta_n)$, we trivially
have the bound $\widetilde{R}(\btheta_n,P) \leq R(\btheta_n,P)$.
Moreover, $\pi_g^n,\bmu_g^n,\bSigma_g^n,b_g^{0,n},\allowbreak
\bb_g^n,\sigma_g^{2,n}$ are convergent sequences when $g\leq H$ and,
then, we can also find a strictly positive constant $R_H$ satisfying
$ 0< R_H \leq \widetilde{R}(\btheta_n,P) \leq R(\btheta_n,P).$ The
sets $B_n=\{(\bx,y):\max_{g\leq H} D_g(\bx,y;\btheta_n)\geq R_H\}$
satisfy that $A_n \subseteq B_n$ and all these $B_n$ sets are
uniformly bounded just by taking into account the uniform continuity
of the set functions $\{(\bx,y) \mapsto \max_{g\leq H}
D_g(\bx,y;\btheta_n)\}_{n=1}^{\infty}$ and that the parameters
corresponding to the first $H$ groups in
$\{\btheta_n\}_{n=1}^{\infty}$ are uniformly bounded.
$\Box$\smallskip

Having established these crucial findings,  we are ready to prove
the existence result.

\subsection*{\bf{Part B: Proof of Proposition 3.2.1}}

Let us start from a sequence $\{\btheta_n\}_{n=1}^{\infty}$
converging toward the supremum of $L(\btheta,P)$. Thanks to Lemma
\ref{l2}, we know that there exists a subsequence of
$\{\btheta_n\}_{n=1}^{\infty}$ with $\bSigma_g^n \rightarrow
\bSigma_g$ and $\sigma_g^{2,n} \rightarrow \sigma_g^2$ for $1\leq g
\leq G$. Moreover, by applying Lemma \ref{l4}, a further subsequence
(with a proper modification, if needed) can be obtained that also
verifies $ \bmu_g^n \rightarrow \bmu_g,
 b_g^{0,n}\rightarrow b_g^0, \bb_g^n \rightarrow \bb_g$ and $\pi_g^n
\rightarrow \pi_g $ with $\pi_g>0$ for any $g$ with $g\leq H$ and
$1< H \leq G$. Let us assume that there exists some $g$ such that
$\bmu_g^n$ is not bounded, or such that  a bounded representation
for $b_g^{0,n}$ and $\bb_g^n$ (in the sense that $\lim\sup_n
E_P\big[ |b_g^{0,n}+(\bb_g^n)' \bX-Y|^2 I_{A_n}(\bX,Y)]=\infty$)
does not exist. We will see that we necessarily must have that
$\pi_g^n \rightarrow 0$ and, consequently, the role played by
$\bmu_g^n, b_g^{0,n}$ and $\bb_g^n$ is irrelevant, given that they
do not modify the value taken by the target function. Therefore, we
could modify them by using other arbitrary convergent parameter
values (of course, satisfying the desired constraints) and the proof
would be done.

To prove that, let us consider
$$
M_n=E_P \left[\left( \log\bigg( \sum_{g=1}^G D_g(\bX,Y;\btheta_n)
\bigg) - \log\bigg( \sum_{g=1}^{H} D_g(\bX,Y;\btheta_n) \bigg)
\right)I_{A_n}(\bX,Y)\right].
$$
By considering the same $R_H>0$ 
used in the proof of Lemma \ref{l4} and the fact that $\log(1+x)\leq
x$, we can see that
$$
M_n \leq \sum_{g=H+1}^{G}E_P \left[ \frac{D_g(\bX,Y;\btheta_n)}{R_H}
I_{A_n}(\bX,Y)\right].
$$
Then, it is trivial to see that $M_n \rightarrow 0$ when $\bmu_g^n$
is not bounded or when no bounded representation for $b_g^{0,n}$ and
$\bb_g^n$ exists for any $g>H$. Consequently, if $\pi_g^n
\rightarrow \pi_g >0$ for any $g>H$ and $\btheta^*$ is the limit of
the subsequence $\{\pi_1^n,...,\pi_H^n, \bmu_1^n,...,\bmu_H^n,
  \bSigma_1^n,...,\bSigma_H^n,\allowbreak  b_1^{0,n},...,b_H^{0,n},  \bb_1^n,..., \bb_H^n, \sigma_1^{2,n}, ...,
  \sigma_H^{2,n}\}_{n=1}^{\infty}$,
we would have that $ \lim_{n\rightarrow \infty} \sup
L(\btheta_n;P)\allowbreak = L(\btheta^*;P)$ (because $M_n
\rightarrow 0$) with $\sum_{j=1}^{H}\pi_j<1$. Then, we could define
a new subsequence
$\{\widetilde{\btheta}_n\}_{n=1}^{\infty}=\{\tilde{\pi}_1^n,...,\tilde{\pi}_G^n,
\tilde{\bmu}_1^n,...,\tilde{\bmu}_G^n,
  \tilde{\bSigma}_1^n,..., \tilde{\bSigma}_G^n, \tilde{b}_1^{0,n},...,\tilde{b}_G^{0,n},
  \tilde{\bb}_1^n,..., \tilde{\bb}_G^n,\allowbreak \tilde{\sigma}_1^{2,n}, ...,
  \tilde{\sigma}_G^{2,n}\}_{n=1}^{\infty}$ with
$$
\widetilde{\pi}^n_{g}=\frac{\pi _{g}^n}{\sum_{g=1}^{k}\pi
_{j}^n}\quad \text{ for } 1\leq g\leq H \quad \text{ and } \quad
\widetilde{\pi}_{H+1}^n=...=\widetilde{\pi}_{G}^n=0,
$$
with $ \widetilde{\bmu}_g^n = \bmu_g^n, \widetilde{b}_g^{0,n} =
b_g^{0,n}, \widetilde{\bb}_g^n = \bb_g^n, \widetilde{\bSigma}_g^n =
\bSigma_g^n \text{ and }\widetilde{\sigma}_g^{2,n} =
\sigma_g^{2,n}\text{ for } 1\leq g\leq H $ and parameters
arbitrarily chosen when $g>H$ (only satisfying the required
constraints). We finally could see that $
 \lim_{n\rightarrow \infty} \sup L(\widetilde{\btheta}_n;P)<
 \lim_{n\rightarrow \infty} \sup L(\btheta_n;P)
$ and this would contradict the optimality stated in the hypothesis
of the present lemma. $\Box$

\subsection*{\bf{Part C: Preliminary results in view of Proposition 3.2.2}}

Before starting the proof of the consistency of the solution for the
sample problem to the population solution, we introduce some
notation, and state some useful results. Let
$\{\hat{\btheta}_n\}_{n=1}^{\infty}=\{\hat{\pi}_1^n,...,\hat{\pi}_G^n,
\hat{\bmu}_1^n,...,\hat{\bmu}_G^n,
\hat{\bSigma}_1^n,...,\hat{\bSigma}_G^n, \hat{b}_1^{0,n},
...,\hat{b}_G^{0,n}, \hat{\bb}_1^n,..., \hat{\bb}_G^n,
\hat{\sigma}_1^{2,n}, ..., \allowbreak
  \hat{\sigma}_G^{2,n}\}_{n=1}^{\infty}\subset \Theta_{c_X,c_{\varepsilon}}$
denote a sequence of empirical estimators obtained by solving the
empirical problems defined from the sequence of empirical measures
$\{P_n\}_{n=1}^{\infty}$.

First, we prove that there exists a compact set $K\subset
\Theta_{c_X,c_{\varepsilon}}$ such that $\hat{\btheta}_n \in K$ with
probability 1. This is done through Lemmas \ref{le4} and \ref{le5},
whose proofs are quite straightforward adaptations of the previously
given proofs of Lemmas \ref{l1}, \ref{l2}, \ref{l3} and \ref{l4}. In
those adaptations, appropriate Glivenko-Cantelli class of functions
must be considered and the class of balls in $\mathbb{R}^{d+1}$
(which is a Glivenko-Cantelli class too)
is taken to provide bounding compact sets when needed.

\begin{lemma}\label{le4}
If $P$ satisfies \emph{(PR)}, then only convergences \emph{(S1)} and
\emph{(V1)} are possible for the $\hat{\bSigma}_g^n$'s and
$\hat{\sigma}_g^{2,n}$'s.
\end{lemma}

\begin{lemma}\label{le5}
If \emph{(PR)} holds, then we can choose a sequence
$\{\hat{\btheta}_n\}_{n=1}^{\infty}$ solving the empirical problem
with components $\hat{\bmu}_g^n$, $\hat{b}_g^{0,n}$ and
$\hat{\bb}_g^n$ such that their norms are uniformly bounded.
\end{lemma}

The following two lemmas are the analogous to Lemmas 5 and 6 in
\cite{GarG:ADAC13}. Their proofs mimic the same steps, with the only
reformulation of the  $D(\cdot;\btheta)$ functions,
 which here take into account the conditional
distribution on the $Y$ variable.

\begin{lemma}\label{le8} Given a compact set $K\subset \Theta_{c_X,c_{\varepsilon}}$,
$B\subset \mathbb{R}^{d+1}$ and $[a,b]\subset \mathbb{R}$, the class
of functions
\begin{equation}\label{gc3}
\mathcal{H}:=\bigg\{
I_B(\cdot)I_{[u,\infty)}\big(D(\cdot,\btheta)\big)\log(
D(\cdot;\btheta)):\btheta \in K,u\in [a,b] \bigg\}
\end{equation} is a Glivenko-Cantelli
class.
\end{lemma}

\begin{lemma}\label{le9} Let $P$ be an absolutely continuous distribution with strictly
positive density function. Then, for every compact set $K$, we have
that
$$
  \sup_{\btheta \in
K}|R(\btheta,P_n) - R(\btheta,P)|\rightarrow 0,\text{ }P\text{-a.e.
.}
$$
\end{lemma}

In fact, the condition on the existence of a strictly positive
density function for $P$ can be removed, but this would imply the
use of trimming functions as those introduced in
\cite{CueG97}.\smallskip

\subsection*{\bf{Part D: Proof of Proposition 3.2.2}}

Taking into account
Lemma \ref{le8}, the consistency 
follows from Corollary 3.2.3 in \cite{van1996weak}, exactly as it
was done in \cite{GarG08} and in \cite{GarG:ADAC13}. Note that
Lemmas \ref{le4} and \ref{le5} guarantee the existence of a compact
set $K$ such that $\{\hat{\btheta}_n\}_{n=1}^{\infty}$ is included
in $K$ with probability 1 and $R(\hat{\btheta}_n,P_n)$ is also
included with probability 1 within an interval $[a,b]$ due to Lemma
\ref{le9}. This has been also used to simplify the target function
needed to apply the aforementioned result in \cite{van1996weak}.
$\square$


\end{document}